\documentclass[fleqn,usenatbib]{mnras}

\usepackage{float}
\usepackage{placeins}
\usepackage{newtxtext,newtxmath}
\usepackage[T1]{fontenc}

\DeclareRobustCommand{\VAN}[3]{#2}
\let\VANthebibliography\thebibliography
\def\thebibliography{\DeclareRobustCommand{\VAN}[3]{##3}\VANthebibliography}

\mathchardef\mhyphen="2D
\newcommand{\tsc}[1]{_{\textsc{#1}}}
\newcommand{\mrm}[1]{{{\mathrm{#1}}}}
\newcommand{\bs}[1]{\boldsymbol{#1}}

\newcommand{\sn}[4]{{S:}{#1}-{V:}{#2}{-}{K:}{#3}-{R:}{#4}}

\newcommand{\rc}{r\tsc{core}}
\newcommand{\tff}{t\tsc{ff}}
\newcommand{\kmin}{k\tsc{min}}
\newcommand{\avir}{\alpha\tsc{vir}}

\newcommand{\kyr}{\mathrm{kyr}}
\newcommand{\yr}{\mathrm{yr}}
\newcommand{\Ms}{\, \mrm{M}_{\odot}}

\newcommand{\pc}{\, \mrm{pc}}
\newcommand{\gcm}{\, \mrm{g} \, \mrm{cm}^{-3}}
\newcommand{\kms}{\mrm{km} \, \mrm{s}^{-1}}
\newcommand{\comment}[1]{}
\newcommand{\vel}{\upsilon}
\newcommand{\tevol}{t\tsc{evol}}
\newcommand{\avg}[1]{\langle #1 \rangle}
\newcommand{\maeg}[2]{\overline{\Delta}_{#1} \, #2}
\newcommand{\mae}[2]{\overline{\Delta}_{\textsc{abs}, #1} \, #2}

\newcommand{\maer}[2]{\overline{\Delta}_{\textsc{rel}, #1} \, #2}

\usepackage{graphicx}	% Including figure files
\usepackage{amsmath}	% Advanced maths commands
\usepackage{mathtools}
\usepackage{xcolor}                                                     
\title[Protostellar Outflows: a window to the past]{Protostellar Outflows: a window to the past}

\author[Rohde et al.]{
P. F. Rohde,$^{1}$\thanks{E-mail: rohde@ph1.uni-koeln.de}
S. Walch,$^{1}$
D. Seifried,$^{1}$
A. P. Whitworth,$^{2}$
and S. D. Clarke$^{1}$ \\
\\
$^1$I. Physikalisches Institut, Universit\"at  zu  K\"oln,  Z\"ulpicher  Str.  77,  D-50937  K\"oln,  Germany\\
$^2$  School  of  Physics  and  Astronomy,  Cardiff  University,  Cardiff  CF24  3AA,  UK\\
}
\date{Accepted XXX. Received YYY; in original form ZZZ}

\pubyear{2021}

\begin{document}
\label{firstpage}
\pagerange{\pageref{firstpage}--\pageref{lastpage}}
\maketitle

%%%%% [Words=245]
\begin{abstract}
During the early phases of low-mass star formation, episodic accretion causes the ejection of high-velocity outflow bullets, which carry a fossil record of the driving protostar's accretion history. We present 44 SPH simulations of $1\Ms$ cores, covering a wide range of initial conditions, and follow the cores for five free-fall times. Individual protostars are represented by sink particles, and the sink particles launch episodic outflows using a subgrid model. The \textsc{Optics} algorithm is used to identify individual episodic bullets within the outflows. The parameters of the overall outflow and the individual bullets are then used to estimate the age and energetics of the outflow, and the accretion events that triggered it; and to evaluate how reliable these estimates are, if observational uncertainties and selection effects (like inclination) are neglected. Of the commonly used methods for estimating outflow ages, it appears that those based on the length and speed of advance of the lobe are the most reliable in the early phases of evolution, and those based on the width of the outflow cavity and the speed of advance are most reliable during the later phases. We describe a new method that is almost as accurate as these methods, and reliable throughout the evolution. In addition we show how the accretion history of the protostar can be accurately reconstructed from the dynamics of the bullets if each lobe contains at least two bullets. The outflows entrain about ten times more mass than originally ejected by the protostar. 
\end{abstract}

\begin{keywords}
methods: numerical -- stars: protostars -- stars: low-mass -- stars: formation -- stars: winds, outflows -- stars: jets
\end{keywords}

%%%%%%%%%%%%%%%%%%%%%%%%%%%%%%%%%%%%%%%%%%%%%%%%%%

%%%%%%%%%%%%%%%%% BODY OF PAPER %%%%%%%%%%%%%%%%%%
%-------------------------------------------------
\section{Introduction}
\label{SEC:Intro}
%-------------------------------------------------

There is growing evidence that accretion onto protostars occurs in episodic bursts rather than being continuous. Accretion events can be observed directly \citep[e.g.][]{Stock20, Rigliaco20, Lee20} or indirectly using chemical modelling \citep[e.g.][]{Hsieh19, Anderl20, Rab20, Sharma20}. So-called FU Orionis (FUor) stars are observed to undergo short outbursts lasting tens of years during which the accretion rate rises to $\sim10 ^{-4}\,{\rm M_\odot\,yr}^{-1}$, followed by long quiescent phases of $\sim 10^3\,-\,10^4$ years with low accretion rates of $ \sim 10 ^{-7}\,{\rm M}_\odot\,{\rm yr}^{-1}$ \citep{Audard14, Safron15, Feher17, Perez20, Takagi20}. These short outbursts of high accretion naturally mitigate the long-standing "luminosity problem" \citep{KenyonSetal1990,Cesaroni18, Hsieh18, Ibryamov18,Kuffmeier18}. Possible causes of episodic accretion are manifold. They include thermal, gravitational, or magneto-rotational instabilities in the accretion disc \citep{Kuffmeier18, Sharma20, Kadam20} and close encounters in multiple systems \citep{Kuruwita20}.

Protostellar outflows accompany the early phases of star formation \citep{Bally16}, and it is widely believed that the launching of protostellar outflows is directly related to accretion onto protostars \citep{Sicilia-Aguilar20}. The mechanisms underlying the launching are still debated \citep[see, e.g. the reviews of][]{Arce07, Frank14, Lee20rev}, but most proposed mechanisms entail the magneto-centrifugal force converting the gravitational energy of the accreted gas into kinetic energy \citep{Blandford82, Konigl00, Lynden-Bell03, Pudritz07, Machida08, Seifried12}. Jets originating from the accretion disc's innermost part are highly collimated and have high velocities \citep{Reipurth01, Tafalla10, Bjerkeli16, Lee17, Gomez-Ruiz19}, whereas winds launched further out in the accretion disc are less collimated and slower \citep{Lee17, Hirota17, Zhang18, Zhang19}. Simulating numerically the inner ejection regions that produce the high-velocity jet component is still a challenging task \citep[e.g.][]{Machida09, Hennebelle11, Seifried12, Price12, Machida13, Machida14, Bate14, Tomida14, Tomida15, Lewis17, Machida19, Saiki20}. The high spatial and temporal resolution required by such simulations is not easily combined with following the outflows on larger time and spatial scales. When focusing on the interaction of outflows with the stellar environment, this problem can be mitigated by introducing an almost resolution independent sub-grid model to launch the outflows \citep{Nakamura07, Cunningham11, Peters14, Myers14, Federrath14, Offner14, Kuiper15, Offner17, Li18, Rohde18}. Once the ejecta are launched, they carve out a cavity by entraining envelope material. Ejecta and entrained material together form a molecular outflow. Sideways motions of bow-shocks \citep{Tafalla17, Jhan21}, together with the wide-angle wind, cause the cavity wall to widen over time \citep{Arce06, Seale08, Velusamy14}.

If the accretion and ejection of gas are strongly coupled, episodic accretion events can be indirectly detected by the episodic outflows they trigger \citep{Arce07, Vorobyov18, Sicilia-Aguilar20}. In particular, where the rapidly ejected high-velocity gas shocks against the slower gas inside the cavity, it produces high-velocity outflow bullets', a frequently observed characteristic of protostellar outflows \citep{chen16, cheng19, Tychoniec19}. In position-velocity diagrams, these outflow bullets -- in this context often called `Hubble Wedges' -- stand out from the otherwise linear position-velocity relation \citep{Bachiller90, Lada96, Arce01, Tafalla04, Garcia09, Wang14, Rohde18, Nony20}. Once an outflow bullet leaves the dense core and appears at optical wavelengths, it is referred to as an Herbig-Haro object. Herbig-Haro objects often form parsec-scale long chains \citep{Reipurth97a, Reipurth98b, Cortes-Rangel20, Ferrero20, Movsessian21}. The spacing and kinematics of outflow bullets in such a chain should carry a fossil record of the underlying episodic protostellar accretion history \citep{Bally16, Lee20rev}.

There are several methods for estimating the age of a young protostellar object. The most common ones involve the spectral energy distribution (SED) of a protostar \citep{Lada87}. As a protostar evolves and grows, the ratio of the mass of the protostar plus disc to the mass of the envelope is expected to increase, altering the SED. This ratio, $(M\tsc{disk} + M_{\star}) \, / \, M\tsc{env}$, can be used to estimate the stellar age \citep{Young05, Vazzano21}. The age can also be estimated by calculating the slope of the SED between two fixed wavelengths \citep[e.g. $2\,\mu\rm{m}$ and $25\,\mu\rm{m}$;][]{LadaCWilkingB1984}; by calculating the bolometric temperature, $T\tsc{bol}$, i.e. the temperature for which a blackbody spectrum has the same flux-weighted mean frequency as the SED \citep{Myers93, Enoch09}; or by calculating the ratio of bolometric to submillimeter luminosity, $L\tsc{bol} \, / \, L\tsc{submm}$, \citep{Andre93, Young05}. A different approach is to estimate ages from chemical abundances in the stellar envelope  \citep[e.g.][]{Tobin13, Busquet17}.

Alternatively, one may study protostellar outflows. (i) Dynamical ages of outflows and their embedded bullets can be used to estimate protostellar ages indirectly \citep{Zhang05, Downes07, Nony20, li2020}. (ii) The outflow activity is expected to diminish over time, which can be used to estimate the driving protostar's evolutionary stage \citep{Curtis10, Yildiz15, Lee20rev, Podio21}. (iii) As the opening angles of the outflow cavities widen over time, there exists a relation between opening angle and age \citep{Arce06, Seale08, Velusamy14, Hsieh17}. However, studying a sample of seven objects in Lupus, \citet{Vazzano21} show that these different methods do not always agree on an evolutionary sequence.

In this paper, we analyse an ensemble of 44 hydrodynamic simulations of low-mass star formation which include episodic protostellar outflow feedback. The outflows consist of ejected gas and core material. The core material is entrained as the outflow carves a bipolar cavity through the core. These simulations enable us to test whether protostellar outflows are a window on the past, i.e. whether they can be used to determine the evolutionary stage, the age, and the accretion history of the underlying protostar.

The paper is structured as follows. In Section\,\ref{sec:Method}, we describe the computational method, the sub-grid outflow model developed earlier by \citet{Rohde18, Rohde20}, and how we define outflow lobes and extract outflow bullets from these lobes. In Section\,\ref{sec:OutflowProperties}, we present outflow properties, estimate entrainment factors and study different velocity components of the simulated outflows. In Section\,\ref{sec:Discussion}, we estimate stellar ages and evolutionary stages from dynamical time scales, outflow rates and cavity opening angles. In addition, we estimate the mean accretion rates and the episodic accretion rates associated with individual outflow bullets. Limitations of the underlying simulations are discussed in Section\,\ref{sec:Caveats}. In Section\,\ref{sec:Conclusions} we summarise our results.\\\\

%-------------------------------------------------
\section{Method}
\label{sec:Method} 
%-------------------------------------------------

%-------------------------------------------------
\subsection{The {\sc GANDALF} SPH code}
\label{sec:Gandalf} 
%-------------------------------------------------

We perform simulations using the smoothed particle hydrodynamics (SPH) and mesh-less finite volume code \textsc{Gandalf} \citep{Hubber18}. We use the `grad-h' SPH formulation \citep{Springel02} with an M4 smoothing kernel, which with $\eta\tsc{sph}\!=\!1.2$ overlaps $\sim\!58$ neighbours. We invoke hierarchical block time-stepping \citep{Hernquist89}. The maximum number of allowed timestep levels is $N\tsc{lvl}\!=\!9$; hence an SPH particle on the highest level receives $2^{N\tsc{lvl}}\!=\!512$ times more updates than a particle on the lowest level. For the time-integration, we adopt the second-order Leapfrog KDK integration scheme. \textsc{Gandalf} uses the artificial viscosity prescription of \citet{Morris97} and the time-dependent viscosity switch of \citet{Cullen10}.

We compute heating and cooling rates using the approximate radiative heating and cooling method of \cite{Stamatellos07} with the modifications developed by \cite{Lombardi15}. The method uses local SPH particle quantities such as density, temperature and  pressure gradient to estimate a mean optical depth. It accounts for changes in the specific heat due to the ionisation of hydrogen and helium, and the dissociation of molecular hydrogen. It also accounts for changes in the opacity, for example due to ice mantle sublimation.

%-------------------------------------------------
\subsection{Feedback models}
\label{sec:FeedbackModels} 
%-------------------------------------------------

In the simulations, protostars are represented by sink particles \citep{Hubber13}, which use a combination of four different sub-grid models, labelled (i) to (iv). Below we give a brief description of these sub-grid models and how they are combined. For further details see \cite{Rohde18, Rohde20} and references therein.

\begin{itemize}
\item[(i)] Episodic accretion is modelled following \citet{Stamatellos12}. Protostars spend most of the time in a quiescent phase with a low accretion rate, $\dot{M}_{\textsc{bg}}\!=\!10^{-7}  \,\Ms \, \mrm{yr}^{-1}$. These quiescent phases lasting $10^3$ to $10^4\,\rm{yr}$ are interrupted by short ($\sim\!50$ yr) outbursts. \citet{Stamatellos12} assume that a combination of gravitational and magneto-rotational instabilities acts as the trigger for these outbursts \citep{Zhu09, Zhu10}. During an outburst the accretion rate quickly increases to $\dot{M}_{\textsc{ob}}\!\simeq\!5 \times 10^{-4}  \,\Ms \, \mrm{yr}^{-1}$ and afterwards decays back to the quiescent value.

\item[(ii)] We adopt the stellar evolution model of \cite{Offner09}, which uses the energy balance between accretion, gravitational contraction, nuclear burning, ionisation and radiation to predict the radius and luminosity of a protostar. Instead of using the accretion rate onto the sink particle, we make use of the accretion rate predicted by the episodic accretion model (i). As a result, the protostar delivers bursts of high luminosity emulating those of FUor type stars.

\item[(iii)] We capture radiative heating due to protostars by invoking a pseudo background radiation field using the luminosities computed from the stellar evolution model (ii). The temperature is assumed to drop with distance, $d$, from a protostar approximately as $d^{-1/2}$. Due to the episodic nature of accretion bursts from protostars, radiative heating temporarily stabilises protostellar accretion discs \citep{Forgan13} while at the same time allowing for some level of disc fragmentation to occur between bursts \citep{Stamatellos12a, Lomax14, Lomax15, Mercer17}.

\item[(iv)] To simulate episodic outflow feedback, we eject $f\tsc{eject}\!=\!10 \%$ of the accreted mass in bidirectional lobes \citep{Nisini18}. In contrast to most outflow sub-grid models we do not use the direct mass accretion rate onto the sink particle, but rather the accretion rate from the episodic accretion subgrid model (i). Consequently the outflow occurs in bursts, leading to the formation of outflow bullets, as frequently observed \citep[e.g.][]{li2020}. The base velocity of the ejected particles corresponds to the Keplerian velocity at twice the stellar radius (ii). To produce a two component outflow, combining a low-velocity, wide-angle wind \citep{Louvet18, Zhang19, Pascucci20, Lee21} and a high-velocity jet, we modulate the base velocity with the angular distribution function derived by \cite{Matzner99} (see Appendix \ref{app:outflow_launching} for more details). Additionally, we add a rotational velocity component (Eq. 9 in \citet{Rohde20}) to capture the angular momentum carried away by the outflow \citep{Lopez20, deValon20, Tabone20}, and thereby keep the angular momentum of the protostar below break-up. SPH particles are ejected in symmetric groups of four to ensure linear and angular momentum conservation.
\end{itemize}

%-------------------------------------------------
\subsection{Simulation setup}
\label{sec:SimulationSetup}
%-------------------------------------------------

We present an ensemble of 44 simulations with different initial conditions. The simulations are identical to the OF-sample, the subset of simulations with outflow feedback, in \citet{Rohde20}. Each run starts from a dense core with mass $M\tsc{core}\!=\!1 \, \Ms$, temperature $T\!=\!10\,\rm{K}$, and a Bonnor-Ebert density profile. We vary three core parameters (i to iii) between the simulations.

\begin{enumerate}

\item The core radius is set to $r\tsc{core}\!=\!0.017\,\pc$, $0.013\,\pc$ or $0.010\,\pc$. The runs with smaller $r\tsc{core}$ are more overcritical, accordingly 3, 4 and 5 times compared to a Bonnor-Ebert sphere in equilibrium. Smaller $r\tsc{core}$ leads to denser cores and correspondingly to shorter free-fall times, respectively, $t\tsc{ff}\!=\!36.8\,\kyr$, $\;24.6\,\kyr\,$ and $\,16.6\,\kyr$. We choose these small core radii to be able to apply a high level of turbulence such that the turbulence is not mostly dissipated before the actual star formation sets in. The dense cores are embedded in a low-density envelope with radius $r\tsc{env}\!=\!0.75$ pc. Outside the core boundary, $r\tsc{core}$, the density decreases as $\rho\propto r^{-4}$ until it falls to $\rho\tsc{env}\!=\!10^{-23} \gcm$. Outside this, the density is uniform. We choose the $r^{-4}$ as a tradeoff between a computationally more expensive shallower profile and a discontinuity. The total envelope mass is $M\tsc{env}\!=\!0.86 \, \Ms$. A fraction of this envelope mass contributes to the final stellar mass as the envelope falls in.

\item The dense cores start out with an initial turbulent velocity field \citep{Walch10}, with power-spectrum
\begin{eqnarray}
P_{k}&\propto&k^{-4},\hspace{1.0cm}k \in\left[\kmin,\,64\right].
\end{eqnarray}
We vary the turbulent velocity field by adjusting the smallest (most energetic) wavenumber between $\kmin\!=\!1,\;2\;\rm{and}\;3$, with $\kmin\!=\!1$ corresponding to the core radius. This changes the turbulent velocity field from small-scale turbulence with low net angular momentum ($\kmin\!=\!3$) to core-scale motions with -- potentially -- high angular momentum ($\kmin\!=\!1$) \citep{Walch12}.

\item We adjust the virial ratio,
\begin{eqnarray}
\avir&=&\frac{2\,(E\tsc{turb} + E\tsc{therm})}{\left|E\tsc{grav}\right|}
\end{eqnarray}
between $\avir\!=\!0.5,\,1.0,\;2.0\;\rm{and}\;3.0$; $\,\avir$ regulates the strength of the turbulence.

\end{enumerate}

Using the crossing timescale $\tau\tsc{cross}$ = 2 $\rc$ / $\upsilon_{\textsc{turb}}$ as an indicator of the dissipation timescale of the turbulence \citep{Stone98, MacLow98}, we find that the dissipation timescales range between $\tau\tsc{diss}$ = 14.0 kyr and 83.4 kyr. These dissipation timescales correspond to free-fall times between  0.8 $\tff$ and 2.3 $\tff$. The first stars in our simulations form between 6.8 and 129.2 kyr with a mean of 27.2 $\pm$ 22.7 kyr, which indicates that the dissipation timescale is comparable to the timescale of the first protostar formation. 

We perform one run for each combination of the three core parameters ($\rc,\kmin,\avir$). For the run with $\avir\!=\!1.0,\,\kmin\!=\!1,\rc\!=\!0.013\,\pc$, we perform eight additional runs with different turbulent random seeds, $\chi$. In total we perform 44 simulations with a mass resolution of $400,000$ SPH particles per $\Ms$ (hence mass resolution $\sim\!3\times 10^{-4}\,\Ms$). The sink creation threshold is $\rho\tsc{crit}\!=\!10^{-10} \, \gcm$. A conservative estimate of the spatial resolution is $4h\tsc{min}\!\sim\!1.2\,\rm{AU}$. The parameters of the individual runs are listed in Table\,\ref{Table:Sims}. For a more detailed description of the simulation setup, and a discussion of the influence of the initial conditions on the simulation outcomes, see \citet{Rohde20}.

\begin{table}
\caption{Parameter summary for all the simulations performed. Reading from left to right, the columns give the run number, the run name, the turbulent random seed ($\chi$), the virial ratio ($\avir$), the smallest turbulent wavenumber ($\kmin$), and the core radius ($\rc /\pc$).} 
\begin{center}
\begin{tabular}{clccccc}
\hline 
\# & Run & $\chi$ & $\avir$ & $\kmin$ & $\rc$ & $\upsilon_{\textsc{turb}}$\\
&&&&& pc & $\kms$
\\
\hline 
1   & \sn{1}{0.5}{1}{17}   & 5 & 0.5 & 1 & 0.017 & 0.4 \\
2   & \sn{1}{1.0}{1}{17}   & 5 & 1.0 & 1 & 0.017 & 0.6 \\
3   & \sn{1}{2.0}{1}{17}   & 5 & 2.0 & 1 & 0.017 & 0.9 \\
4   & \sn{1}{3.0}{1}{17}   & 5 & 3.0 & 1 & 0.017 & 1.1 \\
5   & \sn{1}{0.5}{2}{17}   & 5 & 0.5 & 2 & 0.017 & 0.4 \\
6   & \sn{1}{1.0}{2}{17}   & 5 & 1.0 & 2 & 0.017 & 0.6 \\
7   & \sn{1}{2.0}{2}{17}   & 5 & 2.0 & 2 & 0.017 & 0.9 \\
8   & \sn{1}{3.0}{2}{17}   & 5 & 3.0 & 2 & 0.017 & 1.1 \\
9   & \sn{1}{0.5}{3}{17}   & 5 & 0.5 & 3 & 0.017 & 0.4 \\
10  & \sn{1}{1.0}{3}{17}   & 5 & 1.0 & 3 & 0.017 & 0.6 \\
11  & \sn{1}{2.0}{3}{17}   & 5 & 2.0 & 3 & 0.017 & 0.9 \\
12  & \sn{1}{3.0}{3}{17}   & 5 & 3.0 & 3 & 0.017 & 1.1 \\
13  & \sn{1}{0.5}{1}{13}   & 5 & 0.5 & 1 & 0.013 & 0.5 \\
14  & \sn{1}{1.0}{1}{13}   & 5 & 1.0 & 1 & 0.013 & 0.7 \\
15  & \sn{1}{2.0}{1}{13}   & 5 & 2.0 & 1 & 0.013 & 1.0 \\
16  & \sn{1}{3.0}{1}{13}   & 5 & 3.0 & 1 & 0.013 & 1.3 \\
17  & \sn{1}{0.5}{2}{13}   & 5 & 0.5 & 2 & 0.013 & 0.5 \\
18  & \sn{1}{1.0}{2}{13}   & 5 & 1.0 & 2 & 0.013 & 0.7 \\
19  & \sn{1}{2.0}{2}{13}   & 5 & 2.0 & 2 & 0.013 & 1.0 \\
20  & \sn{1}{3.0}{2}{13}   & 5 & 3.0 & 2 & 0.013 & 1.3 \\
21  & \sn{1}{0.5}{3}{13}   & 5 & 0.5 & 3 & 0.013 & 0.5 \\
22  & \sn{1}{1.0}{3}{13}   & 5 & 1.0 & 3 & 0.013 & 0.7 \\
23  & \sn{1}{2.0}{3}{13}   & 5 & 2.0 & 3 & 0.013 & 1.0 \\
24  & \sn{1}{3.0}{3}{13}   & 5 & 3.0 & 3 & 0.013 & 1.3 \\
25  & \sn{1}{0.5}{1}{10}   & 5 & 0.5 & 1 & 0.010 & 0.6 \\
26  & \sn{1}{1.0}{1}{10}   & 5 & 1.0 & 1 & 0.010 & 0.8 \\
27  & \sn{1}{2.0}{1}{10}   & 5 & 2.0 & 1 & 0.010 & 1.2 \\
28  & \sn{1}{3.0}{1}{10}   & 5 & 3.0 & 1 & 0.010 & 1.4 \\
29  & \sn{1}{0.5}{2}{10}   & 5 & 0.5 & 2 & 0.010 & 0.6 \\
30  & \sn{1}{1.0}{2}{10}   & 5 & 1.0 & 2 & 0.010 & 0.8 \\
31  & \sn{1}{2.0}{2}{10}   & 5 & 2.0 & 2 & 0.010 & 1.2 \\
32  & \sn{1}{3.0}{2}{10}   & 5 & 3.0 & 2 & 0.010 & 1.4 \\
33  & \sn{1}{0.5}{3}{10}   & 5 & 0.5 & 3 & 0.010 & 0.6 \\
34  & \sn{1}{1.0}{3}{10}   & 5 & 1.0 & 3 & 0.010 & 0.8 \\
35  & \sn{1}{2.0}{3}{10}   & 5 & 2.0 & 3 & 0.010 & 1.2 \\
36  & \sn{1}{3.0}{3}{10}   & 5 & 3.0 & 3 & 0.010 & 1.4 \\
37  & \sn{2}{1.0}{1}{13}   & 0 & 1.0 & 1 & 0.013 & 0.7 \\
38  & \sn{3}{1.0}{1}{13}   & 1 & 1.0 & 1 & 0.013 & 0.7 \\
39  & \sn{4}{1.0}{1}{13}   & 2 & 1.0 & 1 & 0.013 & 0.7 \\
40  & \sn{5}{1.0}{1}{13}   & 3 & 1.0 & 1 & 0.013 & 0.7 \\
41  & \sn{6}{1.0}{1}{13}   & 4 & 1.0 & 1 & 0.013 & 0.7 \\
42  & \sn{7}{1.0}{1}{13}   & 6 & 1.0 & 1 & 0.013 & 0.7 \\
43  & \sn{8}{1.0}{1}{13}   & 7 & 1.0 & 1 & 0.013 & 0.7 \\
44  & \sn{9}{1.0}{1}{13}   & 8 & 1.0 & 1 & 0.013 & 0.7 \\
\hline
\end{tabular}\end{center} 
\label{Table:Sims}
\end{table}

%-------------------------------------------------
\subsubsection{Outflow directions}
\label{sec:OutflowDirections}
%-------------------------------------------------
The direction in which outflows are launched depends strongly on the local environment in which the protostar forms, such as the angular momentum axis of the accretion disc and the local magnetic field \citep{Machida20}. Angular momentum of the gas falling onto the accretion disc can significantly alter the orientation of the stellar accretion disc \citep{Matsumoto17}. Therefore, the outflow direction might change significantly over time, possibly causing a quadrupolar outflow as e.g. in \citet{Machida20}. Protostellar companions can cause a precessing jet with a fixed period, as e.g. in \citet{Murphy21}. If multiple protostars form, e.g. via turbulent fragmentation, their outflow directions are random \citep{Lee16},  which can lead to misaligned outflows \citep{Hara21} that may even collide \citep[as described in][]{Zapata18}.

These processes cause a rather chaotic outflow behaviour. However, numerous observations report very ordered outflows, where the outflow axis is straight on parsec scales \citep[e.g.][]{Bally19}. For theoretical astrophysics, it would be of great interest to have some observationally informed statistics of complex versus straight outflows as a benchmark for the simulations. Additionally, such an analysis might allow observers to estimate the level of initial turbulence inside a core by observing the outflows.

\begin{figure*}
    \centering
	\includegraphics[width=0.99\linewidth]{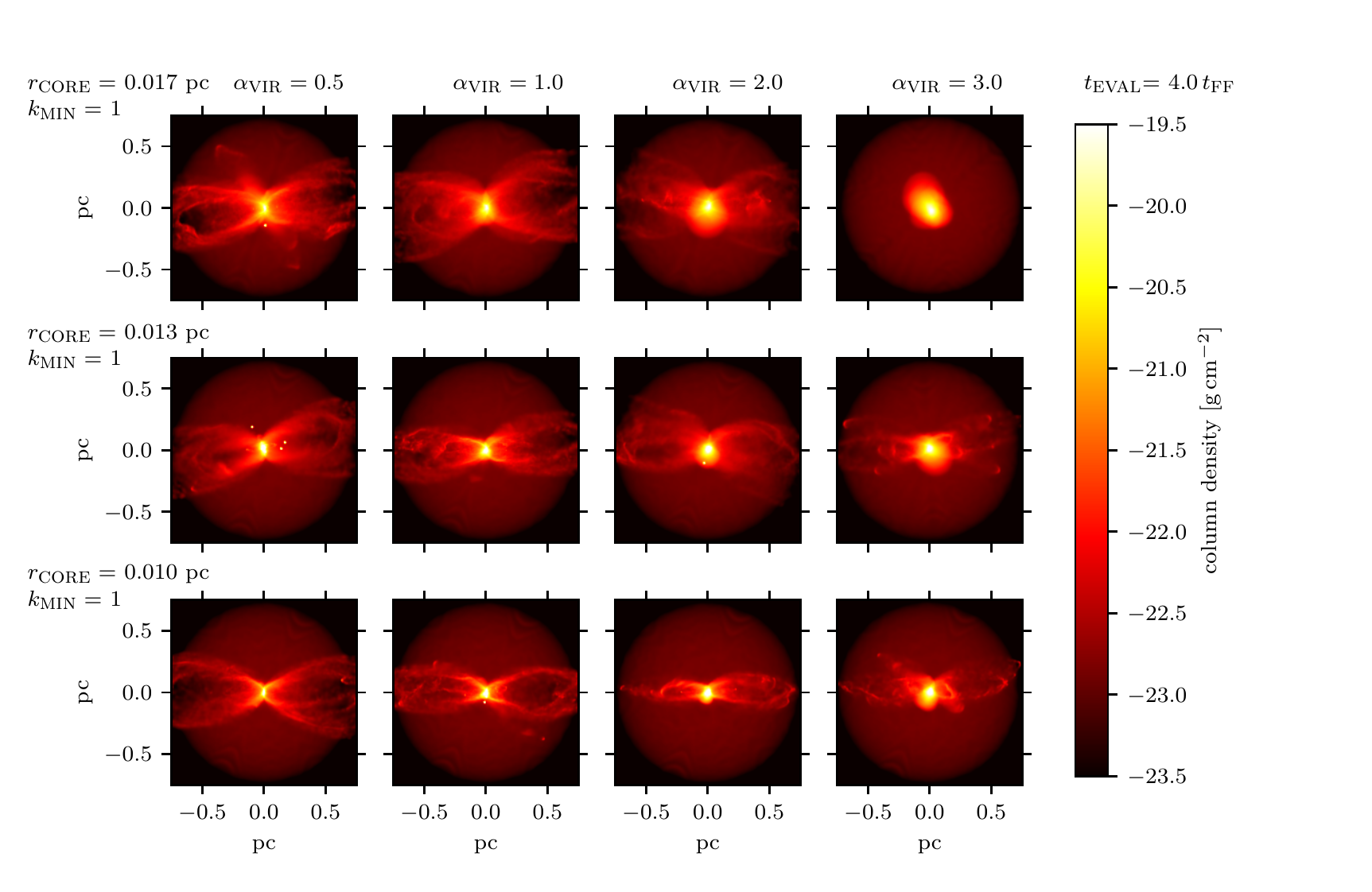}
    \caption{Column density plots of the simulations with $\kmin = 1$ (see Table \ref{Table:Sims}) at $t_{\textsc{eval}} = 4\,\tff$. The rows show simulations with different $r\tsc{core}$, whereas the columns show simulations with different $\avir$. The simulations are rotated such that the prominent outflow lobe preferentially lies along the x-axis. The simulation in the top right panel does not collapse, due to the high level of turbulence in combination with the relatively large core radius.} 
    \label{fig:ColDens1}
\end{figure*}

Fig. \ref{fig:ColDens1} shows column density plots of all simulations with $\kmin = 1$ at $t_{\textsc{eval}} = 4\,\tff$. Some outflows forming in these simulations show a rather complicated behaviour, as e.g. the quadrupolar outflow in the top left panel. From looking by eye there is no clear trend, that simulations with stronger turbulence (higher $\avir$) have a more complicated structure. There is no simulation showing a very straight outflow as e.g. in \citet{Bally19}, which suggests that the general level of turbulence in our simulations could be somewhat too high. Alternatively, the largest mode of the turbulence with respect to the size of the core (here: $\kmin = 1$) could be smaller for real cores, which would lead to a more ordered collapse \citep{Walch12}. Alternatively, including magnetic fields could also stabilize the outflow direction. 

%%%%%
%%%%%

%-------------------------------------------------
\subsection{Outflow-lobe cavity analysis}
\label{sec:CavityAnalyse} 
%-------------------------------------------------

%-------------------------------------------------
\subsubsection{Outflow lobe identification}
\label{sec:Kmean} 
%-------------------------------------------------

In order to analyse the evolution of the outflow kinematics, we need to identify and characterise outflow lobes in SPH simulations. This is not a trivial task. Because of the turbulence in the core, the directions in which outflows are launched change with time, in unpredictable ways. To identify outflows lobes, we use a k-means clustering algorithm \citep[KMCA;][]{Lloyd82}. 

First, we identify all outflowing SPH particles. SPH particle $p$ is tagged as `outflowing' if its speed is higher than the local escape speed, $\vel_{p} > \vel\tsc{esc}(\bs{r}_{p})$, {\it and} $\vel_{p} > 1\, \kms$. Here $\bs{r}_{p}$ is the position vector of particle $p$ with respect to the core's centre of mass (COM), and the escape velocity is given by
\begin{eqnarray} \label{eq:escapeVel} 
\vel\tsc{esc}(\bs{r}_{p})&=&\sqrt{\frac{2GM\tsc{enc}(|\bs{r}_{p}|)}{|\bs{r}_{p}|}} \, ,
\end{eqnarray} 
where $M\tsc{enc}(|\bs{r}_{p}|)$ is the mass enclosed by a sphere around the COM with radius $|\bs{r}_{p}|$. The threshold of $1\,\kms$ is chosen to exclude the weakly bound envelope gas.

Next, we determine $n\tsc{lobe,sim}\,$ k-means clusters, here called lobes, and referenced with the index $l$. Initially, the lobes point in random directions $\bs{e}_{l}$, away from the core's COM. For a specified $n\tsc{lobe,sim}$, we iteratively repeat the following two steps:
\begin{enumerate}
    \item For each SPH particle, we find the lobe with the smallest angle between the particle position $\bs{r}_{p}$ and the lobe direction $\bs{e}_{l}$.
    \item For each lobe, we compute a new lobe direction, $\bs{e}_{l}$, parallel to the mean position vector of all the SPH particles associated with that lobe.  
\end{enumerate}
Iteration ceases as soon as no SPH particle is reassigned to a different lobe in step (i). The number of SPH particles associated with lobe ${l}$ is $n_{\textsc{ass},l}$.

To evaluate whether the resulting lobes represent the structure of the outflow accurately, we use the Silhouette method \citep{Rousseeuw87}, which gives a measure of how well each outflowing particle is represented by the lobe to which it has been assigned, as compared with the neighbouring lobe. The Silhouette for particle $p_{l}$, assigned to lobe $l$, is given by
\begin{eqnarray} \label{eq:silhouette} 
s(p_{l})&=&\frac{b(p_{l}) - a(p_{l})}{\mathrm{max}(b(p_{l}), a(p_{l}))}\,.
\end{eqnarray}
Here $a(p_{l})$ is the mean `distance' between particle $p_{l}$ and all the other particles, $q_{l}$, in lobe $l$, i.e. 
\begin{eqnarray} \label{eq:silhouetteA} 
a(p_{l})&=&\frac{1}{n_{\textsc{ass},l} -1} \sum \limits_{q_{l} = 0, \, q_{l} \neq p_{l}}^{n_{\textsc{ass},l}} \mathrm{dist}\tsc{k-means}(p_{l},q_{l})\,.
\end{eqnarray}
Similarly, $b(p_{l})$ is the mean `distance' between particle $p_{l}$ and all the other particles, $q_{m}$, in the closest neighbouring lobe, $m$, i.e.
\begin{eqnarray} \label{eq:silhouetteB} 
b(p_{l})&=&\frac{1}{n_{\textsc{ass},m}} \sum \limits_{q_{m} = 0}^{n_{\textsc{ass},m}} \mathrm{dist}\tsc{k-means}(p_{l},q_{m}) \, .
\end{eqnarray}
The closest neighbouring lobe is the one that minimises $b(p_{l})$ but does not contain particle $p_{l}$. As a distance measure, $\mathrm{dist}\tsc{k-means}(p,q)$, we use the angle between the two particles $p$ and $q$ with respect to the COM. The Silhouette of a lobe is defined as the mean of all the associated particles' Silhouettes,
\begin{eqnarray} \label{eq:silhouetteC}
S(l)&=&\frac{1}{n_{\textsc{ass},l}} \sum \limits_{p_{l}=0}^{n_{\textsc{ass},l}} s(p_{l}) \, .
\end{eqnarray}
The resulting $S(l) \in [-1,1]$, and the highest value of $S(l)$ corresponds to the best fitting lobe structure. A set containing $n\tsc{lobe,sim}$ lobes is characterised by its mean Silhouette,
\begin{eqnarray} \label{eq:silhouetteMean} 
\bar{S}&=&\frac{1}{n\tsc{lobe,sim}} \sum \limits_{l}^{n\tsc{lobe,sim}} S(l) \, .
\end{eqnarray}

The results of the KMCA depend on the randomly initialised lobe directions $\bs{e}_{l}$. To find the best fitting set of lobes, i.e. the one with the highest $\bar{S}$, we perform the KMCA for ten different sets of initial $\bs{e}_{l}$. One of the ten initial sets is the best fitting one from the previous snapshot.

It is a priori not clear how many lobes are present in a simulation at a given time. Therefore, for each $n\tsc{lobe,sim} \in [2,4,6,8]$ we compute ten KMCA runs; we only consider even numbers of lobes because the sub-grid sink module always launches bipolar outflows. Out of the resulting 40 different realisations, the lobe configuration with the highest mean Silhouette $\bar{S}$ is chosen to represent the outflow cavities. We denote the final lobe vectors as 
\begin{eqnarray} \label{eq:rl} 
\bs{r}_l&=&r_{\textsc{front}, l} \, \bs{e}_l \, ,
\end{eqnarray}
where $r_{\textsc{front}, l}$ is the distance from the COM to the most distant SPH particle allocated to lobe $l$. We repeat this process for each snapshot.

%-------------------------------------------------
\subsubsection{Locating the outflow-lobe cavity wall}
\label{sec:CavityWall} 
%-------------------------------------------------

\begin{figure}
    \centering
	\includegraphics[width=0.8\columnwidth]{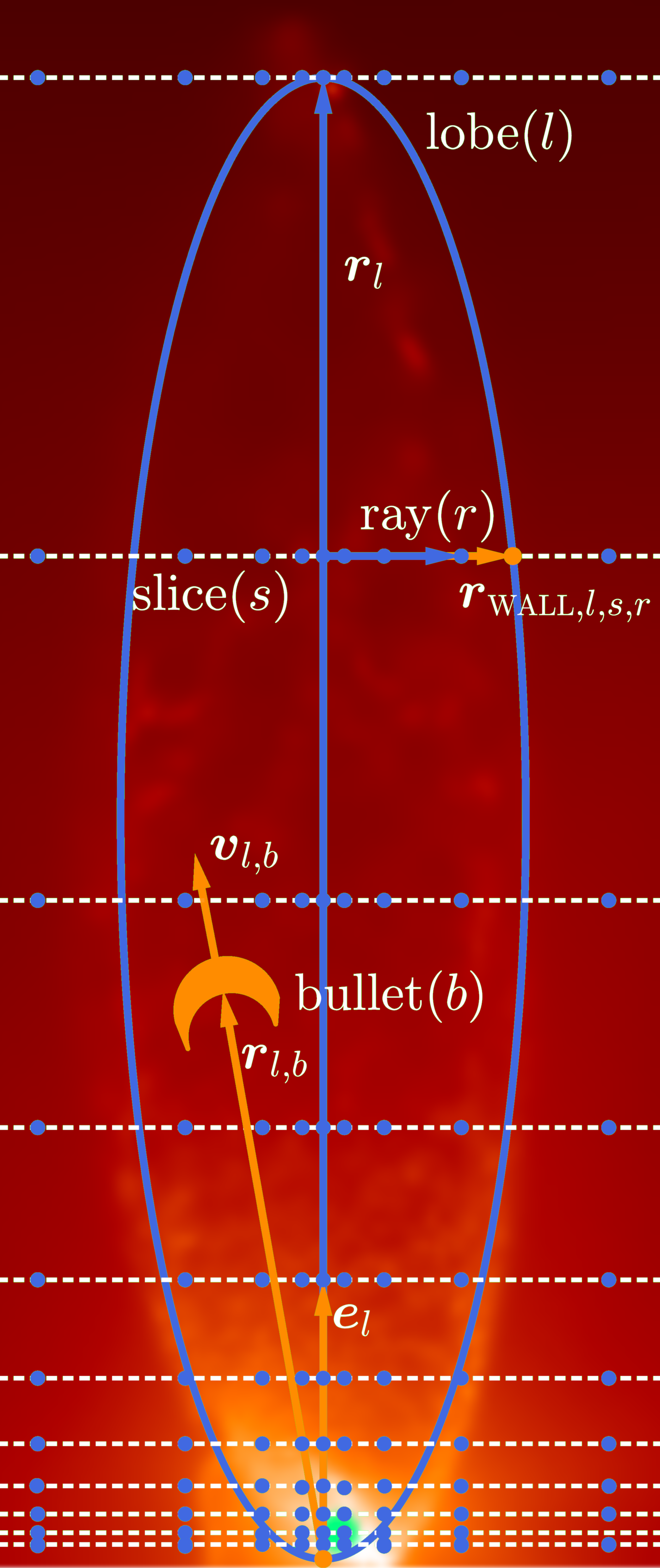}
    \caption{Sketch showing the most important properties of the cavity analysis. The outflow direction, $\bs{e}_{l}$, of lobe $l$, is divided into logarithmically spaced slices (index $s$, dashed white lines) made up of individual rays (index $r$) radiating orthogonally from the lobe axis. Along each ray the simulation is evaluated at logarithmically spaced evaluation points (index $e$, blue dots). Bullets are denoted with the index $b$. The background shows a column density plot from the simulation \sn{5}{3}{3}{13} at $t\!=\!60.6$ kyr.} 
    \label{fig:Sketch}
\end{figure}

To characterise an outflow cavity, we generate an array of evaluation points within each outflow lobe (see Fig. \ref{fig:Sketch}). Along the lobe axis, defined by $\bs{r}_{l}$ (Section\,\ref{sec:Kmean}, Eq.\,\ref{eq:rl}), we place $n\tsc{s}\!=\!64$ perpendicular slices. These slices are logarithmically spaced between $10^{-5}$ and $0.75$ pc (white dashed lines on Fig. \ref{fig:Sketch}), with an additional slice at $10\,\pc$. We use the index $s$ for slices, with $s\!=\!0$ referring to the slice nearest the centre of mass. Each of these slices contains $n\tsc{r}\!=\!90$ rays, $\bs{r}_{l,s,r}$, which are evenly spaced azimuthally around the outflow axis. We use the index $r$ for rays. Each ray consists of $n_{e}\!=\!500$ evaluation points, logarithmically spaced between $10^{-5}$ and $0.5\,\rm{pc}$ radially outward from the lobe axis (blue dots on Fig. \ref{fig:Sketch}). We use the index $e$ for evaluation points. The outer radius of $0.5\,\rm{pc}$ is chosen so that we do not miss any part of a lobe. We use the SPH gather method to evaluate the physical properties of the gas at each evaluation point \citep[see e.g.][]{Monaghan92}.

%%%%%
%%%%%

\begin{figure}
	\includegraphics[width=\columnwidth]{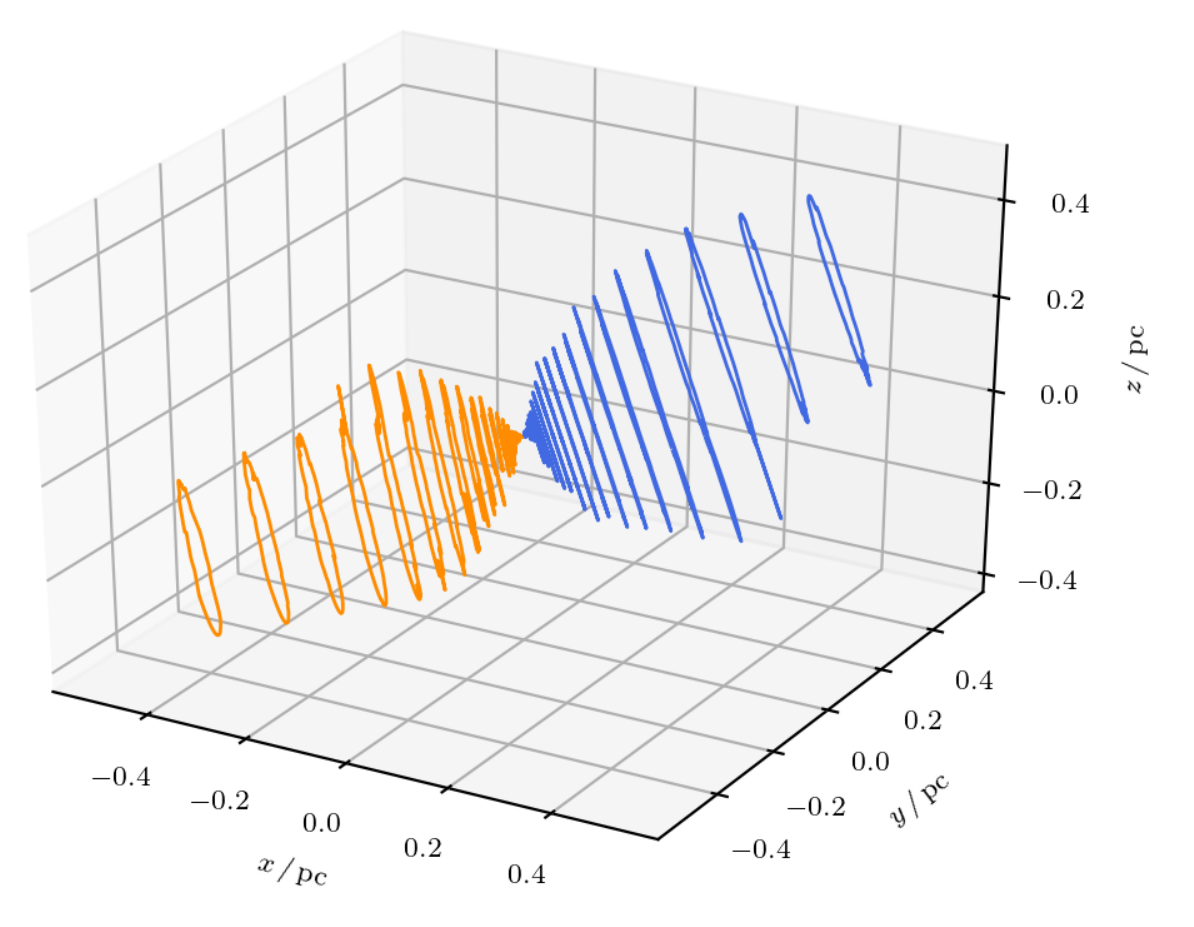}
    \caption{Extracted cavity wall from run \sn{5}{3.0}{3}{13} at $t\!=\!51\,\kyr$. The orange and blue lines lie within evaluation slices perpendicular to the outflow axis and show the extracted cavity wall for both lobes.} 
    \label{fig:Cav3D}
\end{figure}

To define the cavity volume, we walk each ray outwards from the lobe axis. The last evaluation point, $p_i$, for which (a) the gas is outflowing ($\vel_{e}\!>\!\vel\tsc{esc}(\bs{r}_{e})$ and $\vel_{e}\!>\!1\,\kms$), and (b) there is at least one neighbouring particle that is an initially ejected particle, is considered to mark the cavity wall at radius $r_{\textsc{wall},l,s,r}$, provided it is closer to $\bs{r}_{l}$ than to any another outflow axis. With this definition of the cavity wall we find all SPH particles within each lobe, $n_{\textsc{part},l}$. Figure \ref{fig:Cav3D} shows an example of an outflow cavity delineated this way.

%%%%%
%%%%%

%-------------------------------------------------
\subsection{Tracking outflow bullets}
\label{sec:DetBullet} 
%-------------------------------------------------

To locate and track outflow bullets we use the \textsc{Optics} clustering algorithm \citep{Ankerst99}. This algorithm has the advantage that it is density-based and therefore well suited to the Lagrangian nature of SPH simulations. Compared to the frequently used \textsc{Dbscan} algorithm \citep{Ester96}, \textsc{Optics} allows for steeper density gradients, which in our simulations will occur naturally as bullets propagate supersonically through an outflow lobe.

%-------------------------------------------------
\subsubsection{Optics algorithm}
%-------------------------------------------------
\label{sec:Optics} 
Here, we give a brief description of the \textsc{Optics} algorithm, and focus on the modifications needed to apply the algorithm to our simulation data. For a more detailed description, see \cite{Ankerst99}.

The \textsc{Optics} algorithm orders SPH particles in a one-dimensional `reachability list', based on their distance from each other, $\mathrm{dist}\tsc{optics}$. In order to extract high-velocity bullets, we use a distance measure that combines the spatial separation and velocity difference between two particles, $p$ and $q$, viz.
\begin{eqnarray} \label{eq:distMeasureO}
\mathrm{dist}\tsc{optics}(p, q)&=&\sqrt{\left( \frac{ |r_{p}-r_{q}|}{r\tsc{front}} \right)^2 + \left( \frac{ |\vel_{p}-\vel_{q}|}{\vel\tsc{max}} \right)^2} \, 
\end{eqnarray}
Here, $r\tsc{front}$ is the extent of the lobe and $\vel\tsc{max}$ is the highest velocity amongst the particles in the lobe. In order to reduce the computational overhead, we limit application of the \textsc{Optics} algorithm to particles with $\vel_p\!>\!\vel\tsc{min}\!=\!10\,\kms$.

The algorithm involves the following four steps. \\
(i) Starting from a random particle, $p$, we find $p$'s `neighbour list', i.e. the 150 closest particles to $p$ (see Eq. \ref{eq:distMeasureO}) that are also within the same outflow lobe (Section \ref{sec:CavityAnalyse}).\\
(ii) Using this neighbour list we compute the core distance of particle $p$, i.e. the distance from particle $p$ to the $\eta$'th closest particle on $p$'s neighbour list,
\begin{eqnarray} \label{eq:coreDist}
d\tsc{core}(p)&=&\mathrm{dist}\tsc{optics}(p,\eta) \,.
\end{eqnarray}
The number of particles required to form a cluster, $\eta$, is a free parameter, which in the results presented here we have set to $\eta\,=\,15$.\\
(iii) For each particle, $q$, on the neighbour list we compute the reachability distance 
\begin{eqnarray} \label{eq:rdDist}
d\tsc{rd}(p, q)&=&\texttt{max}(\mathrm{dist}\tsc{optics}(p, q), d\tsc{core}(p)) \,, 
\end{eqnarray}
and add this quantity to particle $q$ as a new attribute.\\
(iv) We add particle $p$ to the `reachability list', and the $q$ neighbouring particles, sorted by their reachability distance $d\tsc{rd}$, are added to the `seed list'.

We then take the first particle from the seed list and repeat steps (i) through (iv). In step (iv), if a particle $q$ is already on the seed list and the new $d\tsc{rd}$ is smaller than its former value on the seed list, $q$ is moved forward in the list and $d\tsc{rd}$ is updated. These steps are repeated until all particles in the outflow lobe have been added to the reachability list. Figure \ref{fig:Reachability} illustrates a reachability list.

%-------------------------------------------------
\subsubsection{Extracting bullets}
%-------------------------------------------------

\begin{figure}
	\includegraphics[width=\columnwidth]{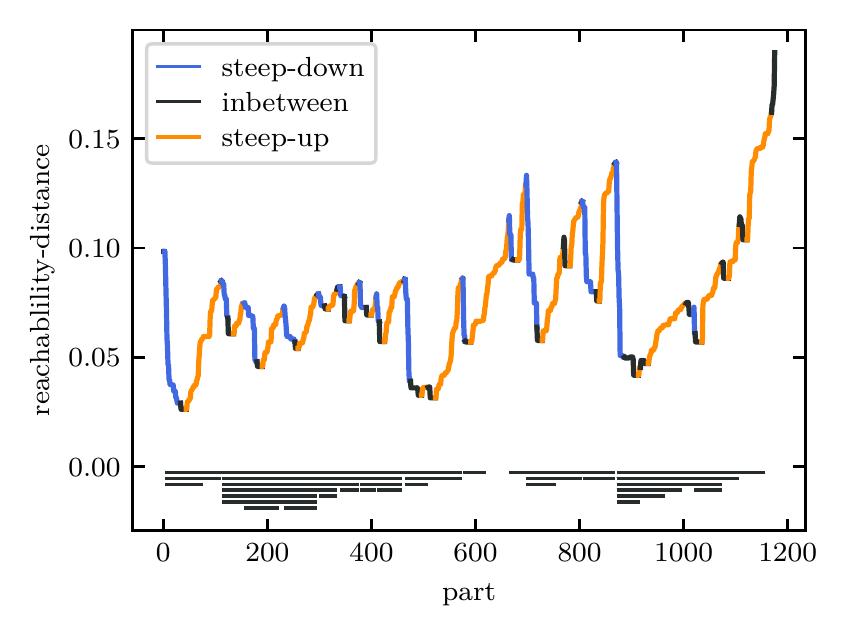}
    \caption{Reachability distance (Eq. \ref{eq:rdDist}) against position in the reachability list for particles from simulation \sn{5}{3.0}{3}{13} at $t\!=\!104\,\kyr$. Following \citep{Ankerst99}, clustered particles are found in dips of the reachability distance. The colour indicates whether a particle is part of a steep-down (blue) or steep-up region (orange). The horizontal black lines on the bottom show the clusters (on the top level) and their sub-clusters.} 
    \label{fig:Reachability}
\end{figure}

The \textsc{Optics} algorithm provides us with an ordered list of the particles' reachability. \cite{Ankerst99} provide an automated method that extracts clusters from the reachability list. Low $d\tsc{rd}(p,q)$ indicates that particles $p$ and $q$ are close, in terms of the chosen distance measure (Eq. \ref{eq:distMeasureO}). On Fig. \ref{fig:Reachability} we plot the reachability distance against the position of the particles on the reachability list. Following \citet{Ankerst99} we identify (sub-)clusters of particles according to `steep-down' (blue) and `steep-up' (orange) regions \citep[plus some other more arcane criteria detailed in][]{Ankerst99}. We sort these (sub-)clusters into an hierarchical structure of clusters and sub-clusters. From top to bottom, the horizontal black lines in Fig.\,\ref{fig:Reachability} show the extracted hierarchical structure of clusters (on the top level) and sub-clusters below.

\begin{figure}
	\includegraphics[width=\columnwidth]{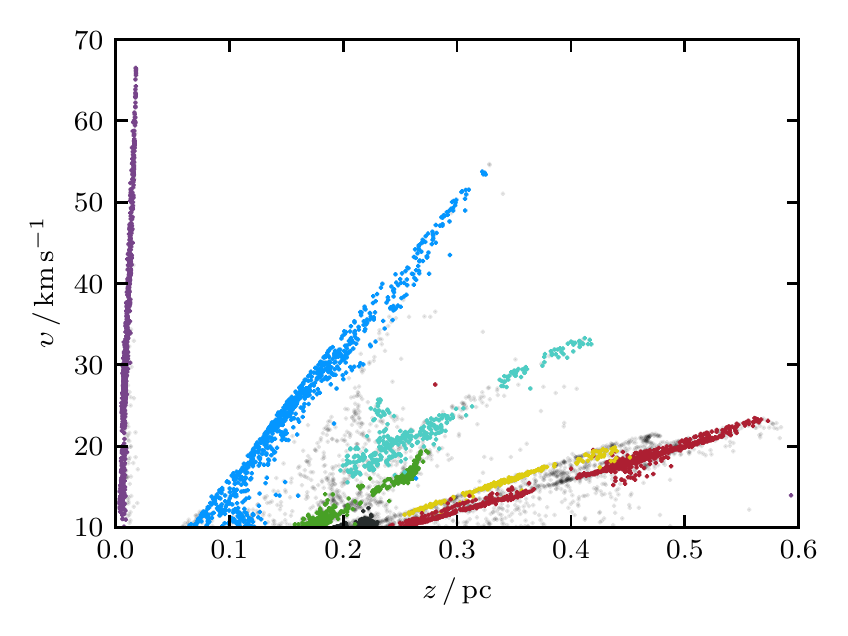}
    \caption{Position-velocity diagram for one of the outflow lobs from simulation \sn{5}{3.0}{3}{13} at $ t\,=\,70 \, \kyr$. Dots represent individual SPH particles. Different colours represent six different outflow bullets identified by the \textsc{Optics} algorithm. Linear features represent individual outflow bullets, called Hubble wedges \citep{Rohde18}.} 
    \label{fig:PV}
\end{figure}

To link a (sub-)cluster to an outburst event, we make use of the subset of particles in that (sub-)cluster that were initially ejected by the protostar. If, from this subset of initially ejected particles, more than $80\,\%$ belong to the same outburst event, we link the (sub-)cluster to that outburst event. We go through the hierarchical structure from top (clusters) to bottom (smallest sub-clusters) until this criterion is fulfilled. Figure \ref{fig:PV} shows the position-velocity diagram of the particles in an outflow lobe, where the colours represent the individual bullets. Since bullets that have decelerated too much are challenging to track, we only consider bullets which contain particles with velocities exceeding 20 $\kms$. The \textsc{Optics} algorithm is very effective in tracing the particles, $n_{\textsc{part},l,b}$, corresponding to a specific outflow bullet, $b$, over the course of a simulation.

%%%%%
%%%%%

%%%%%
%%%%%

%-------------------------------------------------
\section{Outflow properties}
\label{sec:OutflowProperties}
%-------------------------------------------------

We extract all the outflow cavities and bullets from the simulations with episodic outflow feedback in \citet{Rohde20}, using the methodology of Section \ref{sec:Method}. Based on this information we then compute various outflow properties which can be constrained by observation \citep[e.g.][]{Dunham14, Mottram17, li2020}. We note that the observations cover a wide range of core masses from $\sim\!0.1\Ms$ to $\sim\!80 \Ms$, whereas our simulations only treat cores with a mass of $1\,\Ms$. Due to the different initial conditions for the simulations, and their corresponding free-fall times, some cores form protostars faster than others. We evaluate each simulation at time $\tevol\!=t-t_0$, where, $t$ is the simulation time and $t_0$ is the time when the first protostar in a given simulation reaches the threshold mass for protostellar feedback, $M_0\!=\!0.02\,\Ms$.

%To better differentiate between different evolutionary stages, we divide the evolutionary time into a Stage\,0 and Stage\,I \citep{Robitaille06}. A protostar is considered to be in the evolutionary Stage\,0 when the protostellar mass is less than the core mass and transitions to Stage\,I when the protostellar mass exceeds the core mass \citep{Dunham_PPVI}. We stress that there is no one-to-one correlation between the evolutionary Stages and the observational classification scheme (Class 0/I) by \citet{Lada87} based on the infrared spectral index \citep{Dunham_PPVI}. We use as a criterion for Stage\,I that the protostellar mass exceeds the bound core mass. By averaging over all simulations we compute the mean Stage\,0 lifetime as $t\tsc{sI}\,=\,34 \pm 25 \,\kyr$. Our mean Stage\,0 lifetime of $t\tsc{sI}\,=\,34 \, \kyr$ is close to the simulated Stage\,0 lifetimes of 27 kyr by \citet{Dunham12} but significantly lower than observed Class 0 lifetimes of 130 to 260 kyr by \citet{Dunham15}.

The ages of observed protostars cannot be measured directly, but there are several ways in which protostars can be arranged in an approximate evolutionary sequence, as discussed in Section \ref{SEC:Intro}. The most commonly used scheme is the one proposed by   \citet{Lada87}, based on the observed infrared spectral index and the predictions of theoretical models by \citet{Adams86}. This scheme involves three Classes, labelled I, II and III, corresponding to an embedded main accretion phase (Class\,I), a disc accretion phase where the envelope has faded (Class\,II), to an isolated pre-main-sequence phase (Class\,III) \citep{Adams87}. \citet{Andre93} suggest an additional Class\,0 for deeply embedded sources, which show no emission between $2\mu\rm{m}$ and $20\mu\rm{m}$ but have powerful outflows and are defined observationally by their large submillimetre excess, $\,L\tsc{submm}/L\tsc{bol}\!>\!0.005$. However, this classification scheme can be misleading as an evolutionary sequence, because the same object might be classified differently depending on the viewing angle \citep{Calvet94, Crapsi08}. Consequently, \citet{Robitaille06} have proposed a different scheme, involving Stages, and based on the physical properties of the protostar.

As we are interested in evaluating evolutionary stages without explicitly using SEDs, we follow the classification scheme of \citet{Robitaille06}, in which Stage\,0 ends, and Stage\,I begins, when the protostellar mass exceeds the bound core mass \citep{Dunham_PPVI}. Averaged over all the simulations the mean Stage\,0 lifetime as $t\tsc{sI}\!=\!34(\pm 25)\,\kyr$, which is close to the mean simulated Stage\,0 lifetime of $\sim\!27\,\kyr$ reported by \citet{Dunham12}, but significantly lower than observed Class 0 lifetimes of $\sim\!130$ to $\sim\!260\,\kyr$ estimated by \citet{Dunham15}. Reasons why the simulated Stage 0 lifetimes are rather short might be
\begin{itemize}
    \item the small radii of the dense cores, which lead to a rapid collapse,
    \item missing magnetic fields, which would slow down collapse, 
    \item a missing larger scale envelope compared to cores embedded in molecular clouds.
\end{itemize}

%(i) missing magnetic fields, which would slow down collapse, (ii) a missing larger scale envelope compared to cores in molecular clouds, and (iii) the small radii of the dense cores, which lead to a rapid collapse.}

Since we frequently wish to discuss mean values of some quantity, $q_l$, averaged over all the identified lobes from all the simulations, $l\!=\!1$ to $l\!=\!n\tsc{lobe}$, we define
\begin{eqnarray}\label{eq:avg}
\avg{q_l}&=&\frac{1}{n\tsc{lobe}}  \sum_{l=1}^{n\tsc{lobe}} q_{l} \, .
\label{eq:lobaverage}
\end{eqnarray} 
To evaluate the quality of an estimated quantity, $q\tsc{est}$, we compare it with the ground truth, $q\tsc{true}$, for the underlying simulation,  over a specified time interval, $t_0$ to $t_1$. If during this time interval we have $n_s$ snapshots in the range $[n_0,n_1]$, the mean absolute error is
\begin{eqnarray}\label{eq:meanAbsErr}
\mae{[t_0,t_1]}{q\tsc{est}}&=&\frac{1}{n_s} \sum_{i = n_0}^{n_1} |q\tsc{est}(i) - q\tsc{true}(i)|\,,
%\frac{1}{i_{t_1}-{i_{t_0}}} \sum_{i_{t} = i_{t0}}^{i_{t_1}} |q\tsc{est}(i_{t}) - q\tsc{true}(i_{t})|
\end{eqnarray}
and the mean fractional error is
\begin{eqnarray}\label{eq:meanAbsErrRel}
\maer{[t_0,t_1]}{q\tsc{est}}&=&\frac{1}{n_s} \sum_{i = n_0}^{n_1} \frac{|q\tsc{est}(i) - q\tsc{true}(i)|}{\left| q\tsc{true}(i) \right|}\,.
\end{eqnarray}
The important time intervals are Stage\,0 ($t_0\!=\!0\,\rm{kyr}$ to $t_1\!=\!34\,\rm{kyr}$), Stage\,I ($t_0\!=\!34\,\rm{kyr}$ to $t_1\!=\!150\,\rm{kyr}$), and the total simulation time ($t_0\!=\!0\,\rm{kyr}$ to $t_1\!=\!150\,\rm{kyr}$). These time intervals are denoted with indices [s0], [sI] or [s0+I].

%-------------------------------------------------
\subsection{Radius}
%-------------------------------------------------

Given the location of the outflow cavity wall (see Section\,\ref{sec:CavityWall}), we can compute the mean cavity radius (Fig. \ref{fig:Sketch}) for slice $s$ in lobe $l$,
\begin{eqnarray}\label{eq:rSlice}
r_{\textsc{wall},l,s}&=&\frac{1}{n\tsc{r}} \, \sum_{i = 0}^{n\tsc{R}} r_{\textsc{wall},l,s,i} \, . 
\end{eqnarray} 
Here, $r_{\textsc{wall},l,s,i}$ is the distance from the outflow axis to the cavity wall along ray $i$ in slice $s$ of lobe $l$ (Section \ref{sec:CavityWall}). We use the largest mean cavity radius, $\mathrm{max}(r_{\textsc{wall},l,s})$, for all the slices along the outflow axis as the characteristic half-width of the whole outflow lobe, $r_{\textsc{wall},l}$.

\begin{figure}
	\includegraphics[width=\columnwidth]{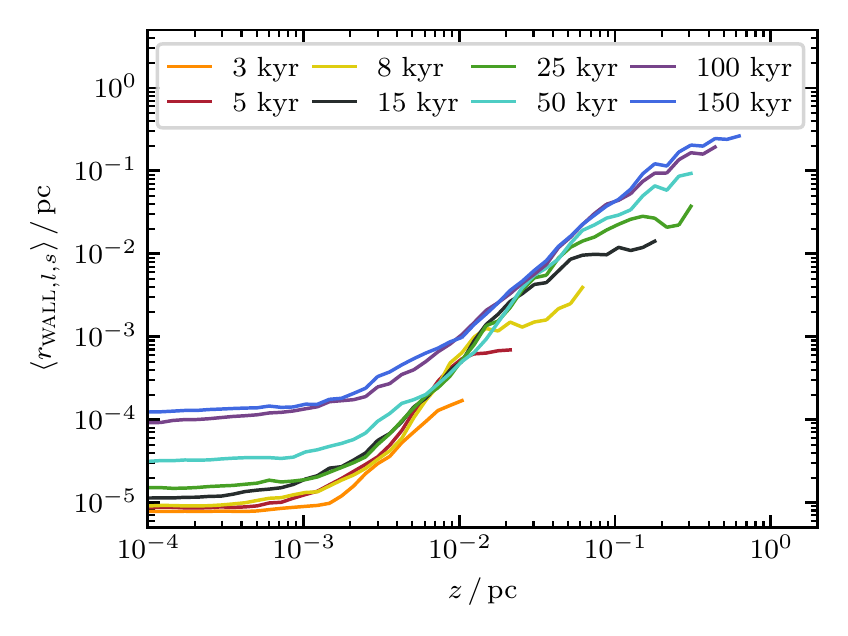}
    \caption{Mean outflow width, $\avg{r_{\textsc{wall},l,s}}$ (Eq. \ref{eq:rSlice}), against the length along the outflow lobe, $z$, at various times $\tevol\,=\,3-150$ kyr (orange to blue). Outflow lobes grow in length and width.} 
    \label{fig:CavWidth}
\end{figure}

Fig.\,\ref{fig:CavWidth} shows the mean cavity width for all the outflow lobes from all the simulations, $\avg{r_{\textsc{wall},l,s}}$, plotted against the length of the outflow lobe, $z$, for eight different times between $t\tsc{evol} = 3$\,kyr and 150\,kyr. Similar to what is observed \citep{Frank14}, the outflow cavity steadily expands with increasing $z$, reaching a maximum width of $0.3\,\rm{pc}$.

%%%%%%%%%%%%%%%%%%%%%%%%%%%%%%%%%%%%%%%%%%%%%%%%%%
%%%%%%%%%%%%%%%%%%%%%%%%%%%%%%%%%%%%%%%%%%%%%%%%%%

%-------------------------------------------------
\subsection{Outflow properties}
\label{sec:ofProp}
%-------------------------------------------------

\begin{figure*}
	\includegraphics[width=\textwidth]{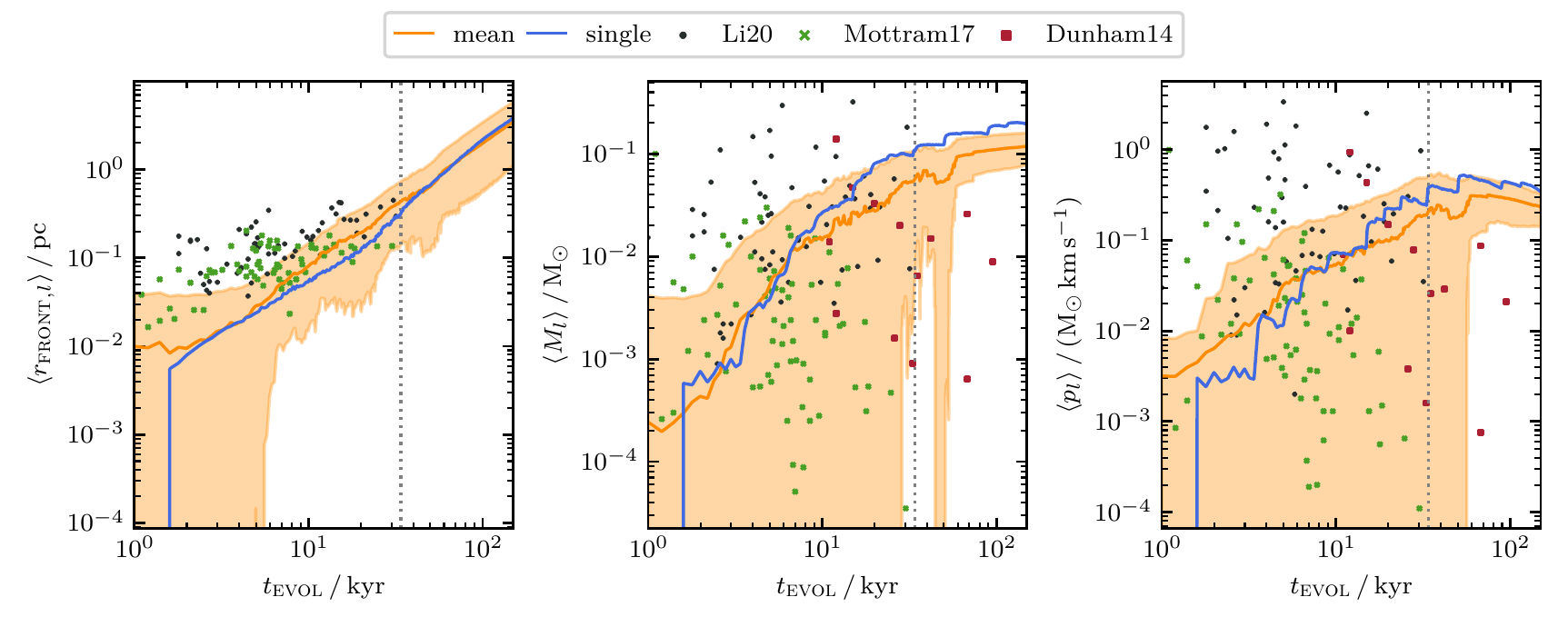}
    \caption{The mean outflow length, $\avg{r_{\textsc{front},l}}$ (left panel), mass, $\avg{M_{l}}$ (middle panel), and momentum, $\avg{p_{l}}$ (right panel), averaged over all the lobes from all the simulations (orange), plotted against the evolutionary time. The orange shaded region shows the standard deviation. For comparison, the blue line shows the results from a single simulation (\sn{7}{1.0}{1}{13}). Observational data from \citet{Dunham14}, \citet{Mottram17} and \citet{li2020} are shown with, respectively, red, green and black markers. \citet{Dunham14} do not provide data for the outflow length. The grey dotted lines indicate the transition from Stage\,0 to Stage\,I averaged over all simulations. The mean outflow length (left) is continuously growing. The mean outflow mass (middle, Eq. \ref{eq:mLobe}) grows asymptotically towards $0.13 \, \Ms$. The mean outflow momentum (right, Eq. \ref{eq:LinMom}) grows during Stage\,0 and decreases during Stage\,I. While the mean mass and mean outflow momentum change quite smoothly, the single simulation shows highly episodic variations.} 
    \label{fig:OverView1}
\end{figure*}

The mean outflow velocity of each lobe $l$ is
\begin{eqnarray}\label{eq:ofVel}
\vel_{l}&=&\frac{1}{n_{\textsc{part},l}} \left| \sum^{n_{\textsc{part},l}}_{p = 0} \bs{\vel}_{l,p} \right|\,,
\end{eqnarray} 
where $\bs{\vel}_{l,p}$ is the velocity of particle p and the summation is over all the particles in the lobe. Similarly, the total momentum of each lobe is
\begin{eqnarray}\label{eq:LinMom}
p_{l}&=&M\tsc{part} \, \left| \sum^{n_{\textsc{part},l}}_{p = 0} \bs{\vel}_{l,p} \right|\,,
\end{eqnarray} 
where $M\tsc{part}$ is the mass of a single SPH particle, and the mass of each lobe is 
\begin{eqnarray}\label{eq:mLobe}
M_{l}&=&n_{\textsc{part},l} \, M\tsc{part} \,.
\end{eqnarray} 
We can then compute means over all the lobes from all the simulations (Eq. \ref{eq:lobaverage}). Fig.\,\ref{fig:OverView1} shows the mean outflow length, $\avg{r_{\textsc{front},l}}$ (left panel; Eq. \ref{eq:rl}), mean outflow mass, $\avg{M_{l}}$ (middle panel; Eq.\,\ref{eq:mLobe}), and mean outflow momentum, $\avg{p_{l}}$ (right panel; Eq.\,\ref{eq:LinMom}), against the evolutionary time (orange lines). The blue line shows these outflow properties for a single simulation (\sn{7}{1.0}{1}{13}). Observational data from \citet{Dunham14}, \citet{Mottram17} and \citet{li2020} are shown with, respectively, red, green and black markers. The time scales on the x-axis for these observations are dynamical ages (see Section \ref{sec:Tdyn}) which depend on the outflow length and should therefore be interpreted as lower limits.

%%%%%
%%%%%

The mean outflow length is almost constant at $\sim\!0.02\,\rm{pc}$ for the first $\sim\!3\,\rm{kyr}$, and thereafter grows continuously with a power-law slope, $\avg{r_{\textsc{front},l}}\!\propto\!t_{\textsc{evol}}^{1.4}$. Our simulations are in good agreement with the observations of \citet{Mottram17}, but somewhat lower than those of \citet{li2020}. The mean outflow mass increases asymptotically towards $\sim\!0.13\Ms$, and is in the same range as in the aforementioned observations during Stage\,0. The mean outflow momentum behaves similarly to the mean outflow mass during Stage\,0. However, the mean outflow momentum has a peak at $\sim\!60\,\rm{kyr}$, and then decreases during Stage\,I. Dividing the mean outflow momentum by the mean outflow mass gives a mean outflow velocity of only a few $\kms$. While the mean outflow mass and momentum are continuous, the single simulation (blue lines) shows strong evidence for episodic outbursts. The extremely low mass and momentum values for some of the lobes (orange shaded region) originate in a few cases where the K-means algorithm finds a lobe, but no corresponding volume, e.g. for inactive ancient lobes. However, this does not mean that the other lobes in such a simulation also have such low outflow masses and momenta.

%-------------------------------------------------
\subsection{Outflow velocity}
\label{sec:ejectionVel}
%-------------------------------------------------

When observing protostellar outflows, the initial ejection velocity is hard to measure, because the ejected gas almost immediately interacts with the cavity or envelope material, and consequently is slowed down. However, to estimate accretion rates or entrainment factors from outflow properties, the ejection velocity is crucial.

\begin{figure}
	\includegraphics[width=\columnwidth]{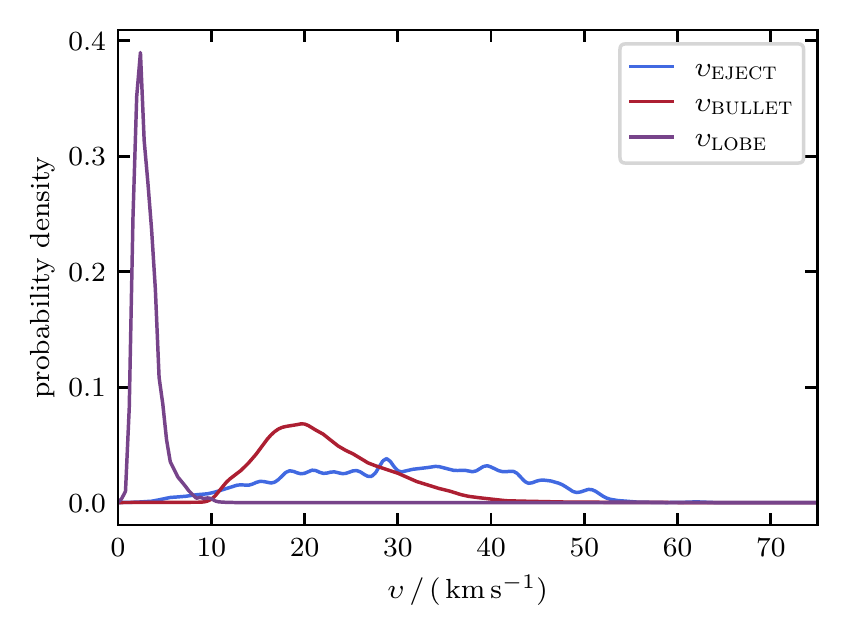}
    \caption{Probability density of the mean outflow- (purple), mean bullet- (red) and mean ejection-velocity (blue) for all lobes from all simulations averaged over $\tevol\!=\!0$ to $150\,\kyr$. The mean outflow velocity peaks at a few $\kms$, and the mean bullet velocity peaks at $\sim 20 \kms$; the mean ejection velocity is more broadly distributed between  $\sim\!10$ and $\sim\!50\,\kms$.}
    \label{fig:OFvelocity}
\end{figure}

Fig.\,\ref{fig:OFvelocity} shows kernel density estimates (KDEs) for three different characteristic velocities, from all the simulated outflows over the period $\tevol\!=\!0$ to $150\,\rm{kyr}$. The purple line shows the mean lobe velocity, $\vel_{l}$ (Eq. \ref{eq:ofVel}), peaking at $\vel\!=\!2.4\,\kms$. The red line shows the mean velocity for all the outflow bullets (Section \ref{sec:DetBullet}). Outflow bullets have higher mean velocities with a peak at $19.6\,\kms$, and maximum values up to $90\,\kms$. Averaging over all outburst events from all simulations gives a mean Keplerian velocity of $15\,\kms$ at the launching radius (i.e. two times the stellar radius). Since we eject particles with this Keplerian velocity modulated by the distribution derived by \citet{Matzner99}, which in our case has a mean value of 2.2 (see Appendix \ref{app:outflow_launching}), the mean ejection velocity is $33\,\kms$. We can estimate the true ejection velocity from the simulations at a given time by computing the mean velocity of all particles ejected during the current time step (blue line). The mean ejection velocities range from $\sim\!10\,\kms$ to $\sim\!50\,\kms$ with a peak at $\sim\!30\,\kms$, in good agreement with the expected ejection velocity. For all further estimates we use ejection velocities between $\vel\tsc{eject}\!=\!20\,\kms$ and $40\,\kms$, with $\vel\tsc{eject}\!=\!30\,\kms$ being the default value.

%%%%%%%%%%%%%%%%%%%%%%%%%%%%%%%%%%%%%%%%%%%%%%%%%%
%%%%%%%%%%%%%%%%%%%%%%%%%%%%%%%%%%%%%%%%%%%%%%%%%%

%-------------------------------------------------
\subsection{Entrainment factor}
%-------------------------------------------------

Ejected outflow gas entrains secondary envelope gas, and together they form a molecular outflow \citep{Tabone17, Zhang19}. The entrainment factor, i.e. the ratio of total outflow mass to ejected mass, is generally not known, but can be estimated if we know the initial ejection velocity.

Assuming that the momentum of the initially ejected gas is conserved, we have  
\begin{eqnarray}\label{eq:EF_2}
M_{l} \, \vel_{l}&=&M_{\textsc{eject},l} \, \vel_{\textsc{eject},l}\,,
\end{eqnarray}
and hence an estimate of the entrainment factor is given by
\begin{eqnarray}\label{eq:EF}
 \epsilon_{\textsc{of},l}\;\;\equiv\;\; \frac{M_{l}}{M_{\textsc{eject},l}}&=&\frac{\vel_{\textsc{eject},l}}{\vel_{l}}\,.
\end{eqnarray} 
Here, $M\tsc{eject}$ is the ejected gas mass, $\vel_{\textsc{eject},l}$ is the ejection velocity (Section\,\ref{sec:ejectionVel}), $\vel_{l}$ and $M_{l}$ are given by Eqs. \ref{eq:ofVel} and \ref{eq:mLobe}. Note that the ejected momentum will actually be somewhat higher than the outflow momentum, since the envelope is still collapsing and therefore contributes momentum opposing the ejected momentum.

\begin{figure}
	\includegraphics[width=\columnwidth]{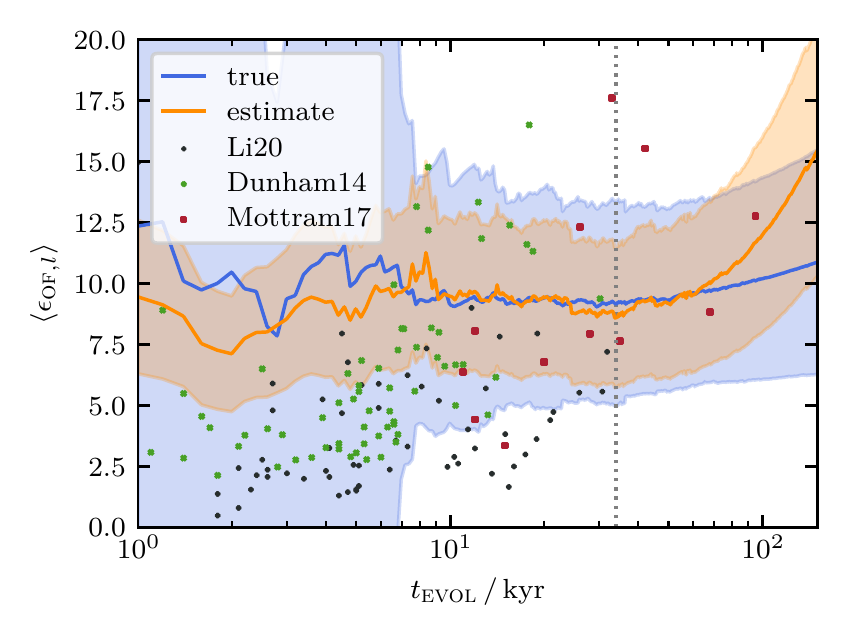}
    \caption{The time evolution of the mean entrainment factors, averaged over all simulations. The blue line shows the mean true entrainment factor, and the blue shading represents the standard deviation. The orange line shows the mean entrainment factor estimated using Eq. \ref{eq:EF} with a fixed ejection velocity of $\vel\tsc{eject}\!=\!30\,\kms$, and orange shading shows the range that is obtained if $\vel\tsc{eject}$ is varied between $20\,\kms$ and $40\,\kms$. Observational data from \citet{Dunham14}, \citet{Mottram17} and \citet{li2020} are shown with -- respectively -- green, red and black markers. The grey dotted line indicates the transition from Stage\,0 to Stage\,I. The mean estimated entrainment factor is close to the mean true entrainment factor during Stage\,0, but at late times tends to give an overestimate.} 
    \label{fig:EntrainmentFaktor}
\end{figure}

Fig.\,\ref{fig:EntrainmentFaktor} compares the time evolution of the true mean entrainment factor (blue line and blue shading for the standard deviation) and the time evolution of the mean entrainment factor estimated using Eq. \ref{eq:EF} with (a) $\vel\tsc{eject}\!=\!30\,\kms$ (orange line) and (b) $20\,\kms\!\leq\!\vel\tsc{eject}\!\leq\!40\,\kms$ (orange shading). For the first $\sim\!7\,\rm{kyr}$, the entrainment factor has a very large spread, but thereafter it settles down to $\avg{\epsilon_{\textsc{of, true},l}}\!\sim\!10(\pm5)$.

For the first $\sim\!\!80\,\rm{kyr}$, the entrainment factor estimated using Eq. \ref{eq:EF} with $\vel\tsc{eject}\!=\!30\,\kms$ resembles the true one very well; the mean absolute error is $\mae{[0 \, \mathrm{kyr}, 80 \, \mathrm{kyr}]}{\avg{\epsilon_{\textsc{of},l}}}\!=\!0.4$. At later times, the mean outflow velocity drops (Fig.\,\ref{fig:OverView1}), causing Eq. \ref{eq:EF} to give an overestimate of the entrainment factor. Over the full time evolution the mean absolute error is $\mae{\textsc{s0+I}}{\avg{\epsilon_{\textsc{of},l}}}\!=\!1.5$. The orange shaded region on Fig.\,\ref{fig:EntrainmentFaktor}  shows the entrainment factor estimated using Eq. \ref{eq:EF} with $20\,\kms\!\leq\!\vel\tsc{eject}\!\leq\!40\,\kms$. This range of ejection velocities translates into a change in the entrainment factor of $\sim\!\pm 4$, which is comparable to the standard deviation of the true entrainment factor.

We have used Eq. \ref{eq:EF} to estimate entrainment factors from the observations of \citet{Dunham14}, \citet{Mottram17} and \citet{li2020}, assuming an ejection velocity of $\vel\tsc{eject}\!=\!30\,\kms$, and compared them with our simulations. The entrainment factors estimated from the \citet{li2020} data (black markers) are about a factor of two smaller, probably because their mean outflow velocity, $\bar{\vel}\tsc{li}\,=\,11.2 \pm 9.2$, is significantly higher than for our simulated outflows. The entrainment factors computed from the  \citet{Dunham14} data (green markers) and the \citet{Mottram17} data (red markers) are correlated with dynamical age. A Kendall Rank Correlation test gives the correlation statistic $\tau\tsc{kr}\!=\!0.51$, with the probability that the null hypothesis is true $p\!\ll\!0.01$. This follows directly from their similarly anti-correlated outflow velocities. Our simulated entrainment factors do not follow this trend.

%%%%%%%%%%%%%%%%%%%%%%%%%%%%%%%%%%%%%%%%%%%%%%%%%%
%%%%%%%%%%%%%%%%%%%%%%%%%%%%%%%%%%%%%%%%%%%%%%%%%%
 
%-------------------------------------------------
\section{Inferring stellar ages and accretion rates from outflows}
\label{sec:Discussion}
%-------------------------------------------------

Observers infer stellar ages and accretion rates from outflow properties \citep[e.g.][]{li2020, Nony20}. Using similar methods, we estimate stellar ages and accretion histories from the outflow properties computed in the preceding section (Section \ref{sec:OutflowProperties}). These estimates are then compared with the underlying simulations to evaluate the different methods. We assume that observational uncertainties and selection effects (like inclination) can be neglected.

%-------------------------------------------------
\subsection{Age estimation}
%-------------------------------------------------

Especially when studying young embedded protostars, it is crucial to have a reliable estimate of the protostellar age. The most common methods for estimating protostellar ages rely on analysing the spectral energy distribution, which provides, for example, the bolometric temperature \citep{Myers93, Enoch09} and the ratio of bolometric to submillimeter luminosity \citep{Andre93, Young05}, and hence a constraint on the protostar's evolutionary stage. Using the spectral index between $2\,\mu\rm{m}$ and $20\,\mu\rm{m}$, sources can be divided into Classes 0 through III \citep{Lada87}, roughly corresponding to evolutionary Stages 0 through III. The drawbacks with these methods are that high angular resolution observations are needed, that the classification depends on the viewing angle \citep{Calvet94, Crapsi08}, and that the distinction between classes constrains the evolutionary stage but not the actual stellar age \citep{Vazzano21}. \citet{Frimann16} perform comprehensive numerical simulations, including radiative transfer modelling, and find that the bolometric temperature and ratio of bolometric to submillimeter luminosity trace the evolutionary stage well but are poor measures of the protostellar age.

%-------------------------------------------------
\subsubsection{Dynamical ages}
\label{sec:Tdyn}
%-------------------------------------------------

An alternative to SED-based methods is to determine dynamical timescales on the basis of the observed properties of outflows. Since outflows occur during the earliest phases of star formation, dynamical ages constrain protostellar ages. Here we compute dynamical ages using five different methods, and evaluate how accurately they reflect true outflow ages, and hence protostellar ages.

The most common method for estimating dynamical ages is to compute the ratio of the lobe extent to the maximum velocity found in the lobe \citep[e.g.][]{Mottram17}
\begin{eqnarray}\label{eq:dyn-velMax}
\tau_{\textsc{max-vel},l}&=&\frac{r_{\textsc{front},l}}{\vel_{\textsc{max},l}} \, .  
\end{eqnarray} 
For this method, $\vel_{\textsc{max},l}$ is defined as the mean velocity of the 50 fastest SPH particles in lobe $l$.

Another common method is to use the terminal speed at the front of the lobe instead of the highest velocity \citep[e.g.][]{Zhang05} 
\begin{eqnarray}\label{eq:dyn-front}
\tau_{\textsc{front},l}&=&\frac{r_{\textsc{front},l}}{\vel_{\textsc{front},l}} \, .   
\end{eqnarray} 
For this method, $\vel_{\textsc{front},l}$ is defined as the mean velocity of the 50 most distant SPH particles in the lobe $l$.

\citet{li2020} estimate the dynamical age using the `perpendicular' method proposed by \citet{Downes07},
\begin{eqnarray}\label{eq:dyn-perp}
\tau_{\textsc{perp},l}&=&\frac{r_{\textsc{wall},l}}{\vel_{l}} \, 
\end{eqnarray} 
where $\vel_{l}$ the mean lobe velocity (Eq. \ref{eq:ofVel}). \citet{Downes07} include an additional factor 1/3 on the righthand side of Eq. \ref{eq:dyn-perp}, to account for inclination uncertainty, but since we neglect inclination, we omit this factor.

In addition to dynamical ages based on the properties of the whole lobe, we can also estimate dynamical ages based on individual outflow bullets. The dynamical age for an individual bullet (Section\,\ref{sec:DetBullet}) is
\begin{eqnarray}\label{eq:dyn-bullet}
\tau_{l,b}&=&\frac{r_{\textsc{max},l,b}}{\vel_{\textsc{max},l,b}}\,,  
\end{eqnarray}
where $r_{\textsc{max},l,b}$ and $\vel_{\textsc{max},l,b}$ are the mean distance and velocity of the bullet's head. $\,r_{\textsc{max},l,b}$ and $\vel_{\textsc{max},l,b}$ are computed by identifying the largest distance and largest velocity among particles in the given bullet and then averaging over all particles in the bullet exceeding 90\% of these largest values. Following \citet{Nony20}, we use the greatest dynamical age among all the bullets in a lobe as an estimate of the lobe's dynamical age, $\tau_{\textsc{bullet},l}$.

Finally, we propose a new method for estimating the dynamical age, which we call the $\Delta$-method. Given two distinct successive outflow bullets in a lobe, we can compute the time scale between the two corresponding outbursts \citep[cf.][]{li2020},
\begin{eqnarray}
\label{eq:dt-bullet}
\Delta t_{l,b}&=&\tau_{l,b+1} - \tau_{l,b}  \, ,
\end{eqnarray} 
where $\tau_{l,b}$ is given by Eq. \ref{eq:dyn-bullet}. If the number of SPH particles in bullet $b$ of lobe $l$ is $n_{\textsc{part},l,b}$, the momentum of bullet $b$ is
\begin{eqnarray}\label{eq:dyn-dBullet_mom}
p_{l,b}&=&M\tsc{part} \, \sum \limits_{p = 1}^{n_{\textsc{part},l,b}} \vel_{l,b,p}\,,
\end{eqnarray} 
and the number of bullets needed to account for the total lobe momentum, multiplied by the time between the two youngest bullets ($b\!=\!0$ and $b\!=\!1$), then gives the dynamical age
\begin{eqnarray}\label{eq:dyn-dBullet}
\tau_{\Delta,l}&=&\frac{2 \, p_l}{p_{l,b=0} + p_{l,b=1}} \,  \Delta t_{l,b=0} \, .
\end{eqnarray} 
We only consider bullets that have $\tau_{\Delta,l,b} > 1 \, \kyr$ since they would otherwise overlap too much.

\begin{figure}
	\includegraphics[width=\columnwidth]{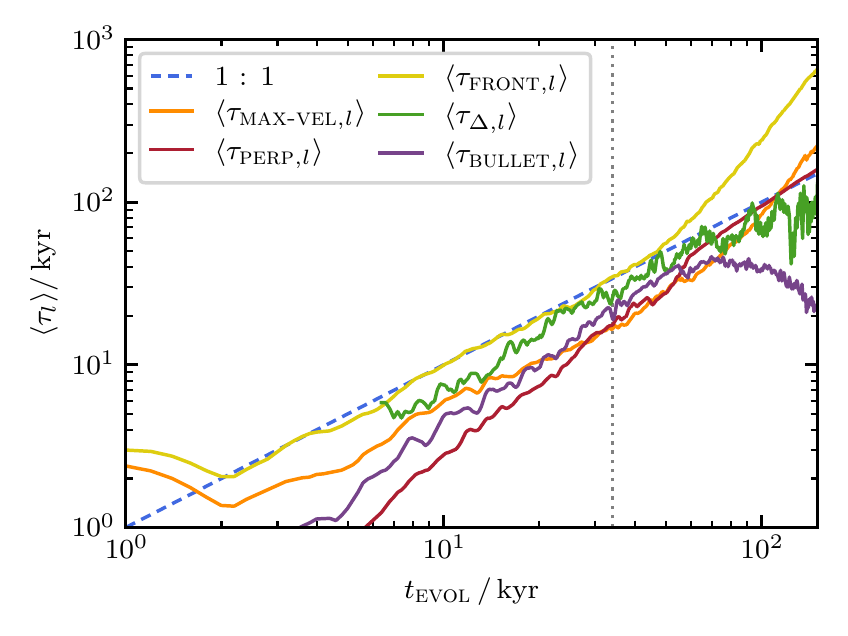}
    \caption{Mean dynamical age estimates, $\avg{\tau_l}$, obtained with the five different methods detailed in Section\,\ref{sec:Tdyn}, plotted against the true protostellar age, for all lobes from all simulations. The blue dashed line shows the one-to-one correlation, and the grey dotted line indicates the transition from Stage\,0 to Stage\,I. The $\tau\tsc{front}$ method gives a good estimate of the protostellar age during Stage\,0, and the $\tau\tsc{perp}$ method gives a good estimate during Stage\,I. The $\Delta$-method gives a good estimate during both Stages, but suffers from being inapplicable at very early times.} 
    \label{fig:tDyn}
\end{figure}

Fig.\,\ref{fig:tDyn} shows the mean dynamical age estimates, $\avg{\tau_l}$, obtained with the five different methods detailed above, plotted against the true protostellar age,  $\tevol$; the dashed blue line indicates one to one correspondence between $\avg{\tau_l}$ and $\tevol$. Table \ref{Table:tDynErr} gives the mean absolute error, $\mae{x}{}$ (Eq. \ref{eq:meanAbsErr}), between $\avg{\tau_l}$ and $\tevol$ averaged over Stage\,0, Stage\,I, and both Stages together; the values in brackets give the mean fractional errors, $\maer{x}{}$ (Eq. \ref{eq:meanAbsErrRel}).

On average, the most commonly used, $\tau_{\tsc{max-vel}}$ method (Eq.\,\ref{eq:dyn-velMax}) underestimates the true protostellar age during Stage\,0 (by $8.4\,\rm{kyr}$; $\maer{\textsc{s0}}{\avg{\tau\tsc{max-vel}}}\!=\!0.47$), and overestimates it during Stage\,I ($\maer{\textsc{sI}}{\avg{\tau\tsc{max-vel}}}\!=\!0.30$).

Using the $\tau_{\tsc{front}}$ method (Eq.\,\ref{eq:dyn-front}) yields a dynamical age, $\tau\tsc{front}$, that is very accurate during Stage\,0, with a mean fractional error of $\maer{\textsc{s0}}{\avg{\tau\tsc{front}}}\!=\!0.08$. However, this method significantly overestimates the protostellar age during Stage\,I, with a mean absolute error of $\sim\!160$ kyr.

The $\tau_{\textsc{perp}}$ method (Eq.\,\ref{eq:dyn-perp}) underestimates the true protostellar age during Stage\,0 ($\maer{\textsc{s0}}{\avg{\tau_{\textsc{perp}}}}\!=\!0.66$) but is significantly more accurate during Stage\,I ($\maer{\textsc{sI}}{\avg{\tau_{\textsc{perp}}}}\!=\!0.15$).

Using the $\tau_\textsc{bullet}$ method (Eq.\,\ref{eq:dyn-bullet}), generally underestimates the protostellar age ($\maer{\textsc{s0+I}}{\avg{\tau\tsc{bullet}}}\!=\!0.55$). This may be because we cannot distinguish bullets that have interacted strongly with the envelope from the rest of the outflow. Observers probably face a similar problem identifying these bullets.

If there are two distinct bullets in the outflow cavity, the $\Delta$-method works well: the estimated dynamical ages, $\tau_{\Delta}$, have the lowest fractional error overall $\maer{\textsc{s0+I}}{\avg{\tau_{\Delta}}}\!=\!0.24$ and the $\Delta$-method performs second best for both Stage\,0 and Stage\,I individually. The high scatter is caused by the reduced sample size of lobes with at least two distinct bullets.

This study ignores observational uncertainties, in particular inclination and selection effects. For an outflow inclined at angle $\theta$ to the line of sight, the outflow velocity is reduced by $\cos(\theta)$, and the length (but not the width) is reduced by $\sin(\theta)$. Therefore, dynamical ages are affected by inclination. For randomly oriented outflows, the mean inclination is $\bar{\theta}\!=\!57.3^{\circ}$ \citep{Bontemps96}, and for this inclination the estimates $\tau_{\textsc{max-vel},l}$ (Eq.\,\ref{eq:dyn-velMax}), $\tau_{\textsc{front},l}$ (Eq.\,\ref{eq:dyn-front}), $\tau_{l,b}$ (Eq.\,\ref{eq:dyn-perp}) and $\tau_{\Delta,l}$ (Eq.\,\ref{eq:dyn-bullet}) will be too high by $\tan(\bar{\theta})\!=\!1.56$, whilst the estimate $\tau_{\textsc{perp},l}$ (Eq.\,\ref{eq:dyn-perp}) will be too high by $\sec(\bar{\theta})\!=\!1.85$. We cannot compute mean correction factors for a random distribution of inclinations, because the integrals involved diverge. In Appendix A we give correction factors for extreme inclinations. However, we note that for low inclinations lengths are very inaccurate, and for high inclinations velocities are very inaccurate.

\citet{Curtis10} and \citet{Vazzano21} argue that dynamical ages are an imprecise measure of protostellar age and may only represent lower limits. Our simulations show that dynamical ages have an average intrinsic error of at least 15\% during Stage\,I in the ideal case of nearly perfect information. Dynamical ages computed from observations might come with significantly larger errors, e.g. due to undetected bullets further out.  If there are two young and distinct bullets in the outflow, the $\Delta$-method seems to be a good alternative to other commonly used methods. Otherwise, we recommend using the perpendicular method since this is the most accurate method during Stage\,I. If it is known from other indicators that the system is still in Stage\,0, the $\tau\tsc{front}$ method is probably the best method, but it is very inaccurate for more evolved systems. The $\tau\tsc{max-vel}$ method gives reasonable estimates overall, whilst the $\tau\tsc{bullet}$ method is the most inaccurate.

%%%%%%%%%%%%%%%%%%%%%%%%%%%%%%%%%%%%%%%%%%%%%%%%%%
%%%%%%%%%%%%%%%%%%%%%%%%%%%%%%%%%%%%%%%%%%%%%%%%%%

%%%%%
\begin{table}
\caption{The mean absolute error (Eq. \ref{eq:meanAbsErr}) in kyr, and the mean fractional error (in brackets, Eq. \ref{eq:meanAbsErrRel}), for the different dynamical age estimates, averaged over Stage\,0, Stage\,I, and both Stages combined.} 
\begin{center}\begin{tabular}{lcccccc}
\hline   
 Method & \multicolumn{2}{c}{$\maeg{\textsc{s0}}{\avg{\tau}}$ / kyr} & \multicolumn{2}{c}{$\maeg{\textsc{sI}}{\avg{\tau}}$ / kyr} & \multicolumn{2}{c}{$\maeg{\textsc{s0+I}}{\avg{\tau}}$ / kyr} \\
\hline 
$\tau\tsc{max-vel}$  & 8.4 & (0.47) & 25.5  & (0.30) & 21.7  & (0.34) \\
$\tau\tsc{front}$    & 0.8 & (0.08) & 159.7 & (1.36) & 125.5 & (1.07) \\
$\tau\tsc{perp}$     & 10.3 & (0.66) & 9.2   & (0.15) & 9.4   & (0.26) \\
$\tau\tsc{bullet}$   & 8.0 & (0.52) & 57.0    & (0.55) & 46.2  & (0.55) \\ 
$\tau_{\Delta}$      & 4.4 & (0.22) & 25.3  & (0.24) & 20.5  & (0.24) \\
\hline
\end{tabular}\end{center} 
\label{Table:tDynErr}
\end{table}
%%%%%

%-------------------------------------------------
\subsubsection{Outflow rates}
\label{sec:ofRates}
%-------------------------------------------------

Another indicator of the protostellar evolutionary Stage is the outflow activity. Outflow activity is expected to be high during Stage\,0 and to decay thereafter \citep{Sperling21}. \citet{Curtis10} and \citet{Yildiz15} find that the outflow rates for momentum and energy are higher in Class\,0 sources than in Class\,I sources. Observers estimate the outflow rates for mass, momentum and energy outflow rates from the outflow properties (Section\,\ref{sec:ofProp}) and the dynamical age (Section\,\ref{sec:Tdyn}). In the sequel we compute outflow rates for our simulated outflows using the same methodology as \citet{li2020} and compare the results obtained when adopting different dynamical age estimates, as well as the true protostellar age.

The mass outflow rate is given by 
\begin{eqnarray}\label{eq:dmdt_out}
\dot{M}_{\textsc{out},l}&=&\frac{M_{l}}{\tau_{l}}\,,
\end{eqnarray} 
the outflow force by
\begin{eqnarray}\label{eq:Force}
F_{l}&=&\frac{p_{l}}{\tau_{l}}\,,
\end{eqnarray} 
and the outflow mechanical luminosity by
\begin{eqnarray}\label{eq:Energy}
L_{l}&=&\frac{1}{2} \, \frac{M_{l} \, \vel_{l}^2}{\tau_{l}}\,.
\end{eqnarray} 

\begin{figure*}
	\includegraphics[width=\textwidth]{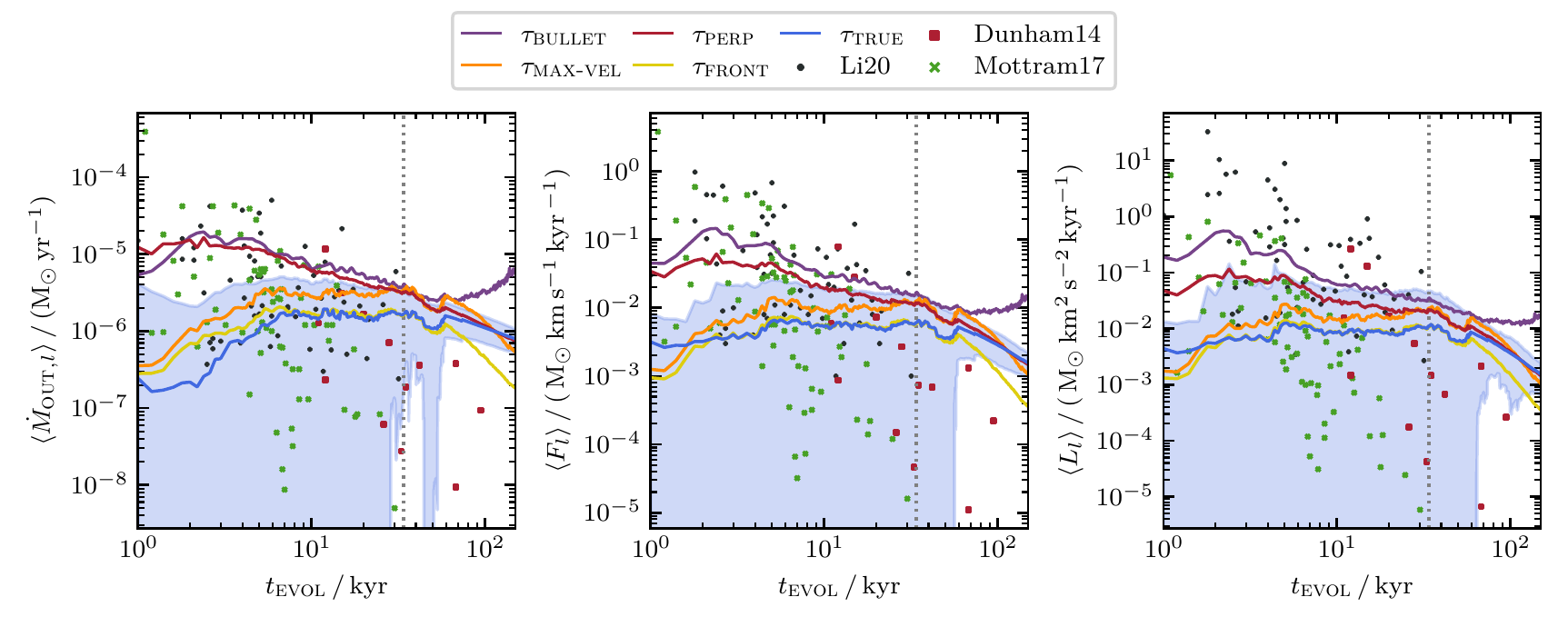}
    \caption{The mean outflow rates for mass (left panel, Eq. \ref{eq:dmdt_out}), momentum (middle panel, Eq. \ref{eq:Force}) and energy (right panel, Eq. \ref{eq:Energy}) computed using different estimated dynamical ages and the true protostellar age, as per the colour code shown in the key; for the true prostellar age, the blue shading shows the standard deviation. The grey dotted vertical lines indicates the transition from Stage\,0 to Stage\,I. Observational data from \citet{Dunham14}, \citet{Mottram17} and \citet{li2020} are shown with -- respectively -- red, green and black marker; the outflow rates for mass and energy from the \citet{Dunham14} observational data are computed using Eqs. \ref{eq:dmdt_out} and \ref{eq:Energy}. The outflow rates for momentum and energy are rather constant in Stage\,0, and slowly decay in Stage\,I.} 
    \label{fig:OverView2}
\end{figure*}

Fig.\,\ref{fig:OverView2} shows the time evolution of the mean outflow rates computed using the different dynamical age estimates derived in Section\,\ref{sec:Tdyn}, and using the true protostellar age (blue line). The outflow rates for mass (left panel), momentum (middle panel) and energy (right panel) show similar trends, increasing for the first $\sim\!5\,\rm{kyr}$, thereafter staying relatively constant during Stage\,0, and then starting to decrease during Stage\,I.

%%%%%%%%%%%%%%%%%%%%%%%%%%%%%%%%%%%%%%%%%%%%%%%%%%
%%%%%%%%%%%%%%%%%%%%%%%%%%%%%%%%%%%%%%%%%%%%%%%%%%

The different dynamical age estimates lead to significantly different estimates of the outflow rates. Some methods work better for young outflows, and others work better for more evolved outflows. Table \ref{Table:overWiev} gives the mean absolute errors and mean fractional errors for Stage\,0, Stage\,I, and the two Stages together.

If dynamical age estimates based on the $\tau\tsc{max-vel}$ method are used (orange line) the outflow rates are overestimated for most of the evolution, and particularly during Stage\,0.

Dynamical age estimates based on the $\tau\tsc{front}$ method (yellow line) yield the most accurate outflow rates during Stage\,0, but thereafter underestimate the outflow rates.

Dynamical age estimates based on $\tau\tsc{perp}$ method (red line) give significantly overestimated outflow rates during Stage\,0, but return the most accurate outflow rates for more evolved outflows.

Dynamical age estimates based on the $\tau\tsc{bullet}$ method (purple line) give significantly overestimated outflow rates for both young and evolved outflows; overall these are the least accurate outflow rates.

Fig. \ref{fig:OverView2} does not show outflow rates computed using dynamical age estimates based on the $\tau_{\Delta}$ method (Eq. \ref{eq:dyn-dBullet}), because not all lobes feature two distinct outflow bullets at the same time, and therefore the blue line is not a valid reference for this method. However, we can compute a valid reference using the true dynamical age for those cases where the $\Delta$-method is applicable and compute the corresponding mean absolute error (see Table \ref{Table:overWiev}). During Stage\,0, the $\tau_{\Delta}$ method yields the second best results after the $\tau\tsc{front}$ method. For more evolved objects the $\tau_{\Delta}$ method provides the second best estimate after the $\tau\tsc{perp}$ method.

The estimated outflow rates from the simulations lie roughly in the middle of the range of observed outflows rates reported by \citet{Dunham14}, \citet{Mottram17} and \citet{li2020}. However, the observed outflow rates decline steeply with time, whereas the rates from the simulations are (a) approximately constant and lower than the observed ones during Stage\,0, and (b) higher than the observed ones and only slowly declining during Stage\,I.

\begin{figure*}
	\includegraphics[width=\textwidth]{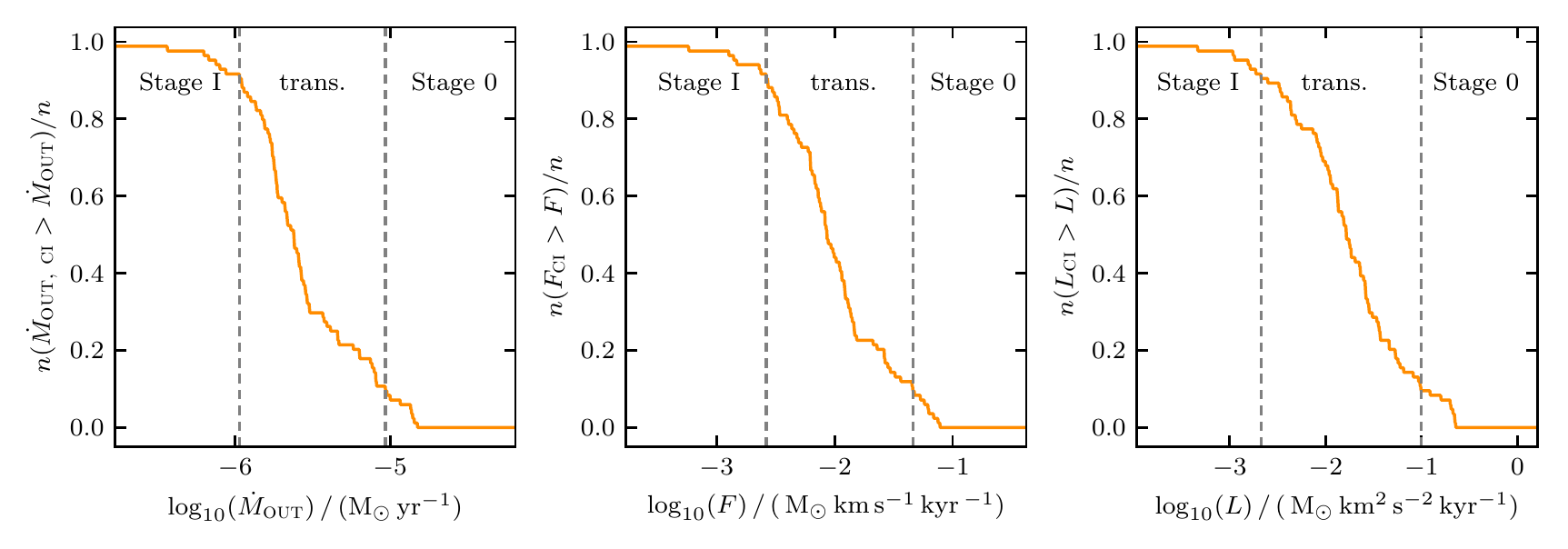}
    \caption{The fraction of lobes that have higher outflow rates for mass (left panel), momentum (middle panel) and energy (right panel) when entering Stage\,I, than the value on the $x$-axis. The vertical grey lines indicate where only 10\% of the lobes have a higher outflow rate (left line) and where only 90\% of the lobes have a lower outflow rate (right line) when entering Stage\,I. Using these thresholds we can define a transition region separating Stage\,0 and Stage\,I sources.} 
    \label{fig:ActivityTransition}
\end{figure*}

Since the core still contains a significant amount of unaccreted mass at the transition between Stage\,0 and Stage\,I, we expect the accretion rate and therefore the outflow activity to diminish slowly across the transition. This slow-down in activity might serve as an indicator of whether the protostar is in Stage\,0 or Stage\,I. In order to quantify the outflow rates for mass ($\dot{M}_{\textsc{out, ci}}$; left panel of Fig. \ref{fig:ActivityTransition}), momentum ($F_{\textsc{ci}}$; middle panel) and energy ($L_{\textsc{ci}}$; right panel) that mark the transition from Stage\,0 or Stage\,I, we compute these rates for each simulation at the time when the protostar enters Stage\,I. In Fig. \ref{fig:ActivityTransition} we count all objects with a transition value larger than the value on the $x$-axis, i.e. we show the fraction of objects that have already transitioned to Stage\,I at a given value of $\dot{M}_{\textsc{out}}$, $F$ or $L$. The objects have a mean gradient of $\,\avg{\nabla \dot{M}_{\textsc{out, ci}}}\!=\!-1.4\times 10^{-7}\,\Ms\,\yr^{-1}\,\kyr^{-1}$, $\,\avg{\nabla F_{\textsc{ci}}}\!=\!-7\times 10^{-4}\,\Ms\,\kms\,\mathrm{kyr}^{-2}$ and $\,\avg{\nabla L_{\textsc{ci}}}\!=\!-2\times 10^{-3}\,\Ms\,\mathrm{km}^2\,\mathrm{s}^{-2}\,\mathrm{kyr}^{-2}$ at the transition to Stage\,I and are therefore mostly decreasing. Hence we interpret the outflow rates on the $x$-axis as an evolutionary sequence, and classify objects as mainly Stage\,0 or Stage\,I on the basis of their respective mass, momentum and energy outflow rates. The vertical grey lines on Fig. \ref{fig:ActivityTransition} indicate the 10$^{\rm th}$ and 90$^{\rm th}$ percentile where most objects are still in Stage\,0 or have already transitioned to Stage\,I, respectively.

The transition region between the 10$^{\rm th}$ and 90$^{\rm th}$ percentile is rather narrow for $\dot{M}_{\textsc{out}}$ and this might facilitate a reliable test of whether the object is in Stage\,0 or Stage\,I. However, our simulation sample is limited to $1\Ms$ cores, and the thresholds might vary significantly with core mass. Distinguishing evolutionary stages becomes even more challenging when account is taken of the fact that outflow rates, especially the energy rate (mechanical luminosity), depend strongly on the inclination angle (see Table. \ref{Table:incl}).

%Using these values of the outflow rates as discriminator we find three region. A region where most of the launching protostars are still in Stage\,0 and a region with mostly Stage\,I sources, separated by a transition region. Therefore, we can classify sources with $\dot{M}_{\textsc{out},l} > 10^{-5.0} \Ms \, \yr^{-1}$, $F_{l} > 10^{-1.3} \Ms \, \kms \, \kyr^{-1}$ or $L_{l} > 10^{-1.0} \Ms \, \mathrm{km}^{2} \, \mathrm{s}^{-2} \, \kyr^{-1}$ as Stage\,I and sources with $\dot{M}_{\textsc{out},l} < 10^{-6.0} \Ms \, \yr^{-1}$, $F_{l} < 10^{-2.6} \Ms \, \kms \, \kyr^{-1}$ or $L_{l} < 10^{-2.7} \Ms \, \mathrm{km}^{2} \, \mathrm{s}^{-2} \, \kyr^{-1}$ as Stage\,0. 

%%%%%%%%%%%%%%%%%%%%%%%%%%%%%%%%%%%%%%%%%%%%%%%%%%
%%%%%%%%%%%%%%%%%%%%%%%%%%%%%%%%%%%%%%%%%%%%%%%%%%

%%%%%%%%%%%%%%%%%%%%%%%%%%%%%%%%%%%%%%%%%%%%%%%%%%
\begin{table*}
\caption{Mean absolute error (Eq. \ref{eq:meanAbsErr}) between mean outflow rates for mass, momentum and energy, estimated using the five different dynamical age estimates detailed in Section\,\ref{sec:Tdyn}, and the corresponding rates obtained using the true protostellar age. The mean absolute errors are given separately for Stage\,0, Stage\,I, and for the whole simulation. The values in brackets are the corresponding mean fractional errors (Eq. \ref{eq:meanAbsErrRel}).} 
\begin{center}\begin{tabular}{lccccccccc}
\hline 
  & \multicolumn{3}{c}{$\maeg{x}{\avg{\dot{M}_{\textsc{out},l}}} \, / \, (10^{-6}\,\Ms \, yr^{-1})$}  & \multicolumn{3}{c}{$\maeg{x}{\avg{F_{l}}} \, / \, (10^{-3} \, \Ms \, \kms \, \kyr^{-1})$} & \multicolumn{3}{c}{$\maeg{x}{\avg{L_{l}}} \, / \, (10^{-3} \, \Ms \, \mrm{km}^{2} \, \mrm{s}^{-2} \, \kyr^{-1})$} \\
 Method & Stage\,0 & Stage\,I & Stage\,0+I & Stage\,0 & Stage\,I & Stage\,0+I  & Stage\,0 & Stage\,I & Stage\,0+I \\
\hline 
$\tau\tsc{max-vel}$  & 
 \,  \ 1.5 \,( 1.15 ) &  \,  \ 0.6 \,( 0.43 ) &  \,  \ 0.8 \,( 0.59 ) &
 \,  \ 4.8 \,( 0.88 ) &  \,  \ 1.8 \,( 0.43 ) &  \,  \ 2.4 \,( 0.53 ) &
 \,  \ 8.1 \,( 0.88 ) &  \,  \ 2.8 \,( 0.43 ) &  \,  \ 3.9 \,( 0.53 ) \\
$\tau\tsc{front}$  & 
 \,  \ 0.2 \,( 0.21 ) &  \,  \ 0.5 \,( 0.47 ) &  \,  \ 0.4 \,( 0.41 ) &
 \,  \ 0.3 \,( 0.06 ) &  \,  \ 1.3 \,( 0.47 ) &  \,  \ 1.1 \,( 0.38 ) &
 \,  \ 0.5 \,( 0.06 ) &  \,  \ 1.7 \,( 0.47 ) &  \,  \ 1.4 \,( 0.38 ) \\
$\tau\tsc{perp}$  & 
 \,  \ 4.4 \,( 6.48 ) &  \,  \ 0.3 \,( 0.22 ) &  \,  \ 1.2 \,( 1.61 ) &
14.3 \,( 3.00 ) &  \,  \ 1.0 \,( 0.22 ) &  \,  \ 3.9 \,( 0.84 ) &
24.1 \,( 3.00 ) &  \,  \ 1.6 \,( 0.22 ) &  \,  \ 6.6 \,( 0.84 ) \\
$\tau\tsc{bullet}$  & 
 \,  \ 5.5 \,( 7.14 ) &  \,  \ 2.4 \,( 2.38 ) &  \,  \ 3.1 \,( 3.44 ) &
27.9 \,( 5.96 ) &  \,  \ 7.2 \,( 2.80 ) & 11.8 \,( 3.51 ) &
75.1 \,( 9.94 ) & 11.2 \,( 3.28 ) & 25.4 \,( 4.77 ) \\ 
$\tau_{\Delta}$  & 
 \,  \ 1.6 \,( 0.32 ) &  \,  \ 0.5 \,( 0.38 ) &  \,  \ 0.8 \,( 0.37 ) &
 \,  \ 8.0 \,( 0.32 ) &  \,  \ 1.7 \,( 0.38 ) &  \,  \ 3.1 \,( 0.37 ) &
19.9 \,( 0.32 ) &  \,  \ 2.9 \,( 0.38 ) &  \,  \ 6.8 \,( 0.37 ) \\
\hline
\end{tabular}\end{center} 
\label{Table:overWiev}
\end{table*}
%%%%%%%%%%%%%%%%%%%%%%%%%%%%%%%%%%%%%%%%%%%%%%%%%%

%-------------------------------------------------
\subsubsection{Opening angle}
\label{sec:openingAngle}
%-------------------------------------------------

Protostellar outflows do not only expand along the outflow direction; they also widen over time \citep{Frank14, Hsieh17}. We compute the time evolution of the opening angles of the simulated outflows, and compare them to the observational data from \citet{Arce06},  \citet{Seale08} and \citet{Velusamy14}, to assess whether opening angles might be used to estimate protostellar ages.

We first compute the opening angle of each slice along the outflow axis (see Section \ref{sec:CavityWall} and Fig.\,\ref{fig:Sketch}),
\begin{eqnarray}\label{eq:openingAnge}
\phi_{\textsc{open},l,s} = 2 \, \arctan{\left( \frac{r_{\textsc{perp},l,s}}{z_{l,s}} \right)} \, .  
\end{eqnarray} 
Then we define the opening angle of the lobe, $\phi_{\textsc{open},l}$, as the largest $\phi_{\textsc{open},l,s}$ along the outflow axis, with the constraint that $z_{l,s}\!>\!0.002\,\rm{pc}$. We use this threshold to exclude extremely large opening angles close to the source.%; varying this threshold value has no significant impact on the results.

\begin{figure}
	\includegraphics[width=\columnwidth]{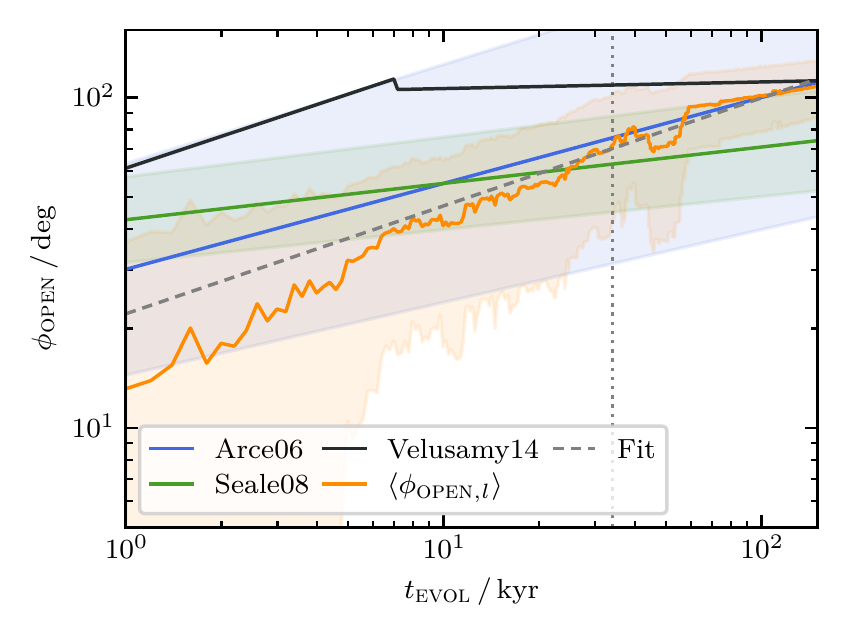}
    \caption{Mean opening angle against protostellar age. The orange line shows the mean opening angle, $\avg{\phi_{\textsc{open},l}}$, with its standard deviation (shaded region) for all outflow lobes from all simulations. The grey dotted vertical line indicates the transition from Stage\,0 to Stage\,I, and the grey dashed line shows the best power-law fit to the simulation results (as per Eq.\,\ref{eq:powerLawFit}). The blue and green lines show the relations derived by -- respectively -- \citet{Arce06} and \citet{Seale08} from observations, with the corresponding shaded regions representing the uncertainties. The black line shows the relation derived by \citet{Velusamy14} from observations. Our results match the results of \citet{Arce06} well, and confirm that the opening angles of outflow cavities widen over time. } 
    \label{fig:openingAngle}
\end{figure}

Fig.\,\ref{fig:openingAngle} shows how the opening angle varies with protostellar age according to \citet{Arce06} (blue line),  \citet{Seale08} (green line) and \citet{Velusamy14} (black line). The orange line shows the mean opening angle for all lobes from all simulations, $\avg{\phi_{\textsc{open},l}}$, together with the corresponding standard deviation (orange shaded region). The grey dashed line shows a least-squares fit to the time evolution of $\avg{\phi_{\textsc{open},l}}$, of the form
\begin{eqnarray}\label{eq:powerLawFit}
\phi\tsc{open}(t) = b \, [t/\rm{kyr}]^{\alpha},
\end{eqnarray} 
with parameters $b\!=\!22.2(\pm0.3)^\circ$ and $\alpha\!=\!0.327(\pm0.003)$. The mean opening angles range from $\avg{\phi_{\textsc{open},l}} \!\sim\!13^\circ$ to $\sim\!110^\circ$, with the trend that more evolved outflows have higher opening angles. A Kendall Rank Correlation test yields a correlation statistic $\tau\tsc{kr}\!=\!0.39$, with a probability that the null hypothesis is true $p\!\ll\!0.01$.

Comparing our simulated outflows with observations, we find significantly lower opening angles during the first $\sim\!7\,\rm{kyr}$. Thereafter, the simulations are in good agreement with the observations of \citet{Arce06}. The power-law relation obtained by \citet{Arce06} has an exponent $\alpha\tsc{arce}\!=\!0.26$, which is close to our fit with $\alpha\!=\!0.327$ and within the standard deviation (orange shaded region). Our relation is significantly steeper than the one found by \citet{Seale08}. The broken power-law relation of \citet{Velusamy14} predicts much larger opening angles during the early evolution, and only matches our results at late times, $\sim\!150\,\rm{kyr}.\,$ \citet{Offner11} use a completely different method to compute the opening angles in their simulations, but their opening angles are in good agreement with ours.

\begin{figure}
	\includegraphics[width=\columnwidth]{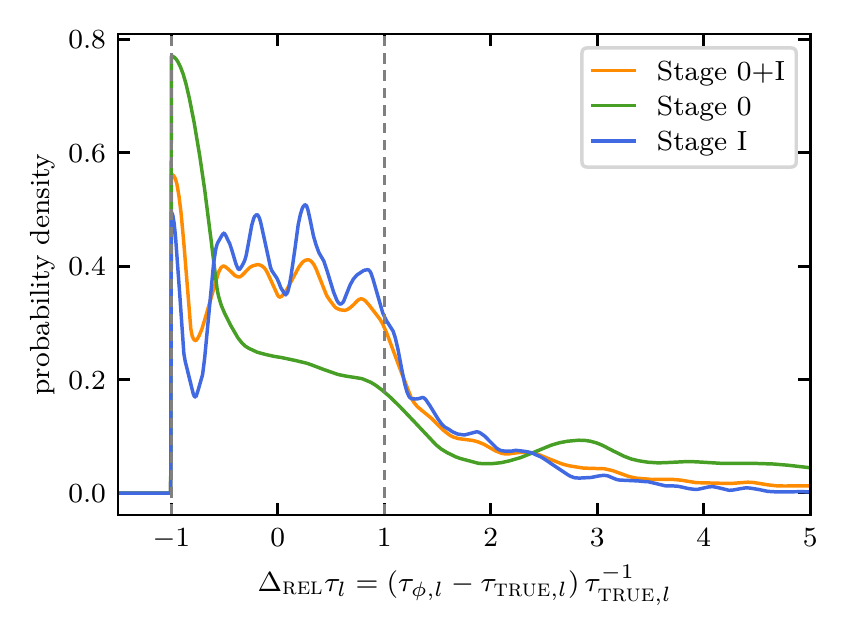}
    \caption{Probability density of the fractional error (i.e. $\Delta\tsc{rel}\!\tau_{\phi}\!=\!(\tau_{\phi}\!-\!\tau_{\textsc{true}})/\tau_{\textsc{true}}$) for all dynamical age estimates, $\tau_{\phi}$, made using the opening angle--age relation (Eq. \ref{eq:powerLawFit}). The green line shows the probability density of the error for Stage\,0, the blue line for Stage\,I, and the orange line for both Stages together. The high chance of seriously over- or under-estimating the true protostellar age demonstrates that the opening angle--age relation is not a reliable method for estimating protostellar ages.}
    \label{fig:openingAngle_error}
\end{figure}

Even though we find a correlation between opening angle and age, the opening angle--age relation gives very imprecise estimates of the protostellar age. Due to turbulence, we find a large scatter in the opening angles at a given age (orange shaded region on Fig. \ref{fig:openingAngle}), which in combination with the shallow slope of $\alpha\!=\!0.327$, translates into large errors in the estimated age; even a  small change in $\phi\tsc{open}$ produces a large change in the estimated age. In Fig. \ref{fig:openingAngle_error} we plot the distribution of fractional errors when using the opening angle--age relation to estimate the age of a simulated outflow. Especially during Stage\,0, we underestimate the protostellar age significantly. There is a prominent peak in the distribution of errors at $\Delta\tsc{rel} \tau_{\phi}\!\sim\!-1$, which corresponds to an underestimate of the true age by almost $100\!\%$. During Stage\,I, the distribution of errors is somewhat better, but there is still a high chance of an error of $\sim\!100\%$. Another reason for the inaccuracy of the estimates is that a power-law fit does not fully describe the time evolution of the opening angles. In the first $\sim 10\,\kyr$, the actual slope is steeper than $\alpha\!=\!0.327$; and, since the opening angles must have a maximum, it starts to flatten at $\sim\!60\,\rm{kyr}$.

\begin{figure}
	\includegraphics[width=\columnwidth]{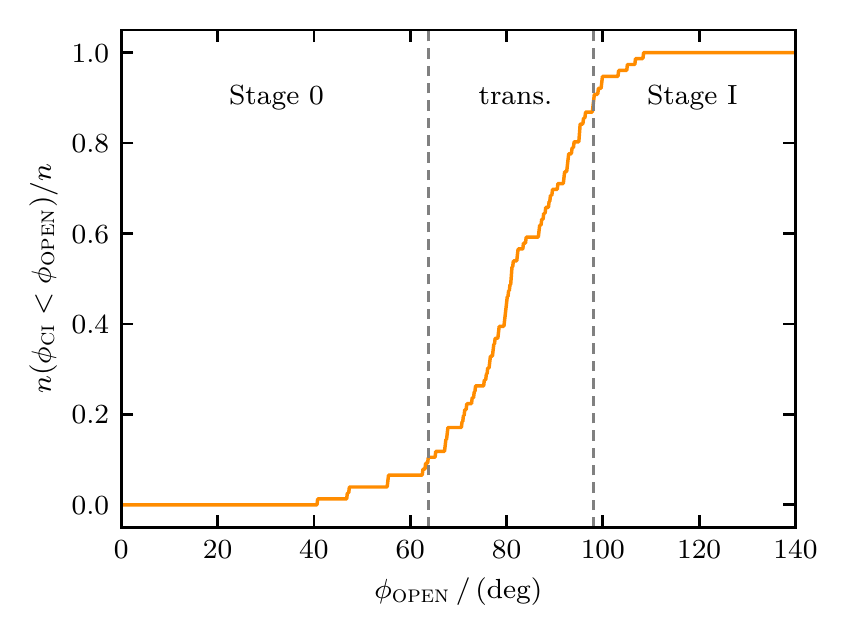}
    \caption{The cumulative distribution of opening angles when the simulated lobes transition to Stage\,I. The dashed grey lines indicate the opening angles below which only $10\%$ of the lobes have a smaller opening angle (left line), and above which only $10\%$ have a larger opening angle (right line), when entering Stage\,I. The opening angles can be divided into three regimes: sources with $\phi_{\textsc{open},l}\!<64\!^{\circ}$ (left line) are very likely to be in Stage\,0, and those with opening angles $\phi_{\textsc{open},l}\!>\! 98^{\circ}$ (right line) are very likely to be in Stage\,I; those with intermediate $\phi_{\textsc{open},l}$ might be in either Stage.} 
    \label{fig:openingAngle_CI_transition}
\end{figure}

Even though the opening angle does not give an accurate estimate of the protostellar age, it still helps us to distinguish between Stage\,0 and Stage\,I. Fig \ref{fig:openingAngle_CI_transition} shows the cumulative distribution of opening angles when the simulated lobes transition to  Stage\,I. Only 10\% of simulations have $\phi\tsc{open}\!<\!64^{\circ}$ when transitioning to Stage\,I, and only 10 \% of simulations have $\phi\tsc{open}\!>\!98^{\circ}$. Thus we conclude that outflows with opening angles $\phi\tsc{open}\lesssim 65^{\circ}$ are very likely to be in Stage\,0, and those with $\phi\tsc{open}\gtrsim 100^{\circ}$ are very likely to be in Stage\,I. In between there is a transition region where both Stages are possible. \citet{Arce06} find a similar division, with most YSOs with $\phi\tsc{open}\!<\!55^{\circ}$ being Class\,0, and most YSOs with $\phi\tsc{open}\!>\!75^{\circ}$ being Class I.

%%%%%%%%%%%%%%%%%%%%%%%%%%%%%%%%%%%%%%%%%%%%%%%%%%
%%%%%%%%%%%%%%%%%%%%%%%%%%%%%%%%%%%%%%%%%%%%%%%%%%

%%%%%%%%%%%%%%%%%%%%%%%%%%%%%%%%%%%%%%%%%%%%%%%%%%
%\section{Accuracy of opening angle--age relation}
%%%%%%%%%%%%%%%%%%%%%%%%%%%%%%%%%%%%%%%%%%%%%%%%%%

%%%%%%%%%%%%%%%%%%%%%%%%%%%%%%%%%%%%%%%%%%%%%%%%%%
%%%%%%%%%%%%%%%%%%%%%%%%%%%%%%%%%%%%%%%%%%%%%%%%%%

%-------------------------------------------------
\subsection{Accretion history}
\label{sec:AccHist}
%-------------------------------------------------

%-------------------------------------------------
\subsubsection{Outflow bullet age}
\label{sec:tDynBullet}
%-------------------------------------------------

Outflow bullets carry valuable information about the accretion history of the launching protostar. Knowing when a bullet was ejected allows us the reconstruct this accretion history. Here, we estimate the ejection times of outflow bullets using their dynamical ages, $\tau_{l,b}$ (Eq. \ref{eq:dyn-bullet}), and compare them to the true ages, $\tau_{\textsc{true},l,b}$. The error is
\begin{eqnarray}\label{eq:bulletTdynErr}
\Delta \tau_{l,b}\,=\,\tau_{l,b} - \tau_{\textsc{true},l,b}\,.
\end{eqnarray}

\begin{figure}
	\includegraphics[width=\columnwidth]{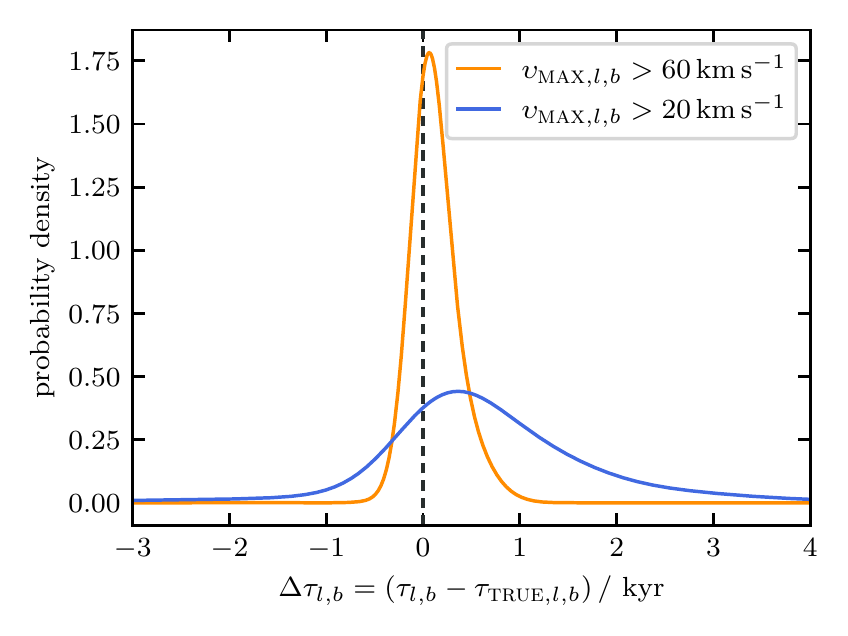}
    \caption{Probability density of the difference between the estimated bullet dynamical age, $\tau_{l,b}$ (Eq. \ref{eq:dyn-bullet}), and the true dynamical age, $\tau_{\textsc{true},l,b}$. The orange line represents only bullets with $\vel_{\textsc{max},l,b}\!>\!60\,\kms$. The blue line represents all bullets with $\vel_{\textsc{max},l,b}\!>\!20\,\kms$. Dynamical ages inferred from high-velocity outflow bullets are very reliable.} 
    \label{fig:BulletTdyn}
\end{figure}

Fig.\,\ref{fig:BulletTdyn} shows the KDEs of $\Delta \tau_{l,b}$ for all bullets with $\vel_{\textsc{max},l,b}\!>\!20\,\kms$ (blue line), and for high-velocity bullets only with $\vel_{\textsc{max},l,b}\!>\!60\,\kms$ (orange line). High-velocity bullets tend to be recently ejected and have not yet been decelerated much by their environment, so they give an accurate record of the accretion history. Their error distribution (orange line on Fig.\,\ref{fig:BulletTdyn}) is highly peaked. A Gaussian fit to KDE shows that the distribution is peaked at $\Delta \tau \!=\!0.07\,\rm{kyr}$ and has a width of $\sigma = 0.12\,\rm{kyr}$. The bullets' ejection times can therefore be estimated very precisely; on average, we overestimate the ejection time by $\lesssim 0.1\,\rm{kyr}$. A Gaussian fit to the error distribution for all bullets with $\vel_{\textsc{max},l,b}\!>\!20\,\kms$ (blue line on Fig. \ref{fig:BulletTdyn}) is much broader ($\sigma\!=\!0.5\,\rm{kyr}$) and the peak is shifted to a higher $\Delta \tau \!=\!0.44$\,\rm{kyr}.

Hence the dynamical age of an outflow bullet gives a good estimate of the ejection time, especially if the bullet is young and has not been significantly decelerated by its environment. However, one should keep in might that the dynamical age is always affected by inclination (see Table \ref{Table:incl}).

%%%%%%%%%%%%%%%%%%%%%%%%%%%%%%%%%%%%%%%%%%%%%%%%%%
%%%%%%%%%%%%%%%%%%%%%%%%%%%%%%%%%%%%%%%%%%%%%%%%%%

%-------------------------------------------------
\subsubsection{Time-averaged accretion rate}
\label{sec:lobeAcc}
%-------------------------------------------------

%The kinematic information carried by the outflow allows us to estimate a mean accretion rate of the driving source. Similar to \citet{li2020}, we compute an ejection rate using the outflow rate (Eq. \ref{eq:dmdt_out}) and the entrainment factor (Eq. \ref{eq:EF}) as
%\begin{eqnarray}\label{eq:ejectRate}
%\dot{M}_{\textsc{eject},l} = \epsilon_{\textsc{of},l}^{-1} \, \dot{M}_{\textsc{out},l} \, .
%\end{eqnarray} 
%We use the true dynamical age (Section\,\ref{sec:Tdyn}) to compute the outflow rate. The ejection rate translates into an accretion rate by accounting for the ejection fraction $f\tsc{eject}$ as
%\begin{eqnarray}\label{eq:accRate}
%\dot{M}_{\textsc{acc},l} = \frac{2}{f\tsc{eject}} \dot{M}_{\textsc{eject},l} \, .
%\end{eqnarray} 
%Here, the additional factor of two compensates that we derive the accretion rate for each lobe individually. Inserting Eq.\,\ref{eq:EF} and Eq.\,\ref{eq:dmdt_out} the accretion rate follows as
%\begin{eqnarray}
%\label{eq:accRateFull}
%\dot{M}_{\textsc{acc},l} = \frac{2}{f\tsc{eject}} \frac{\vel_{l}}{\vel\tsc{eject}} \, \frac{M_{l}}{\tau_{l}} \, 
%\end{eqnarray} 
%and for typical values as
%\begin{eqnarray}
%\label{eq:accRateInsert}
%\begin{aligned}
%\dot{M}_{\textsc{acc},l} = 2 \times 10^{-5} \frac{\Ms}{\yr} \,
%\left(\frac{\vel_{l}}{3 \, \kms} \right) \,
%\left(\frac{M_{l}}{0.01 \, \Ms} \right)
%\\
%\times \,
%\left( \frac{f\tsc{eject}}{0.1} \, \right)^{-1}
%\left( \frac{\vel\tsc{eject}}{30 \, \kms} \right)^{-1} \,
%\left( \frac{\tau_{l}}{\kyr} \right)^{-1} \, .
%\end{aligned}
%\end{eqnarray}

The kinematic information carried by the outflow allows us to estimate the time-averaged accretion rate onto the underlying protostar. The mass ejection rate is given by
\begin{eqnarray}\label{eq:ejectRate}
\dot{M}_{\textsc{eject},l}&=&\frac{\dot{M}_{\textsc{out},l}}{\epsilon_{\textsc{of},l}}
\;\,=\;\,\frac{M_l\,\vel_l}{\vel\tsc{eject}\,\tau_l},
\end{eqnarray}
where the second expression on the righthand side is obtained by substituting for $\dot{M}_{\textsc{out},l}$ from Eq.\,\ref{eq:dmdt_out}, and for $\epsilon_{\textsc{of},l}$ from Eq.\,\ref{eq:EF}. The time-averaged mass accretion rate is then given by
\begin{eqnarray}\label{eq:accRate}
\dot{M}_{\textsc{acc},l}\!\!&\!\!=\!\!&\!\!\frac{2\,\dot{M}_{\textsc{eject},l}}{f\tsc{eject}}\;\,=\;\,\frac{2\,M_l\,v_l}{f\tsc{eject}\,\vel\tsc{eject}\,\tau_l}\\\nonumber
\!\!&\!\!=\!\!&\!\!2 \times 10^{-5} \Ms\,\yr^{-1}\,
\left(\frac{M_{l}}{0.01 \, \Ms} \right)
\left(\frac{\vel_{l}}{3 \, \kms} \right)\,\hspace{1.2cm}\\\label{eq:accRateInsert}
\!\!&\!\!\!\!&\!\!\hspace{0.8cm}\times\,
\left( \frac{f\tsc{eject}}{0.1} \, \right)^{-1}\,
\left( \frac{\vel\tsc{eject}}{30 \, \kms} \right)^{-1}\,
\left( \frac{\tau_{l}}{\kyr} \right)^{-1},
\end{eqnarray}
where the factor 2 on the righthand side of Eq.\,\ref{eq:accRate} derives from the fact that there are two outflow lobes.

\begin{figure}
	\includegraphics[width=\columnwidth]{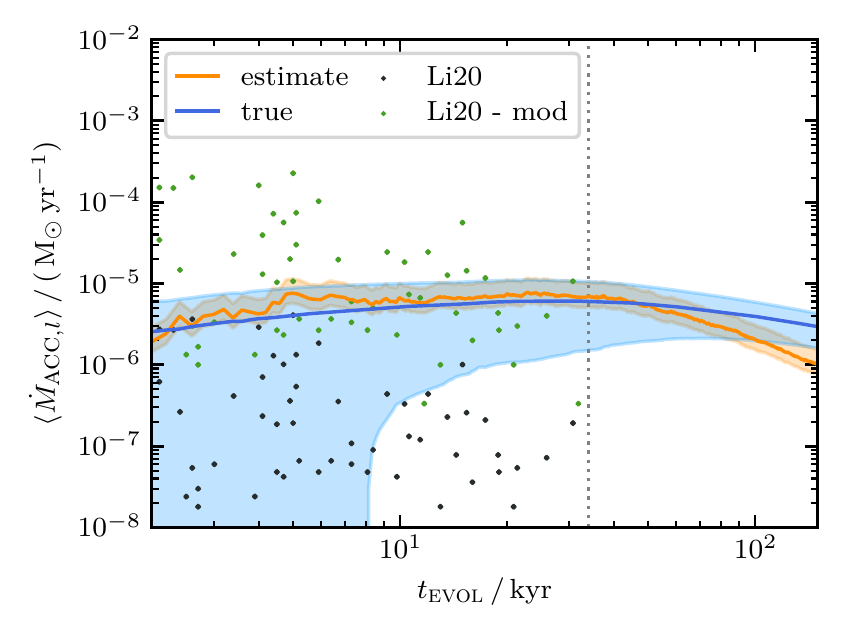}
    \caption{Mean accretion rate, $\avg{\dot{M}_{\textsc{acc},l}}$, against time, $t\tsc{evol}$. The blue line shows the mean {\it true} accretion rate for all lobes from all simulations, with the standard deviation delineated by the blue shaded region. The orange line shows the mean accretion rate estimated using Eq. \ref{eq:accRate} with $\vel\tsc{eject}\!=\!30\,\kms$; and the orange shaded region shows the range of accretion rates if $\vel\tsc{eject}$ is varied between $20\,\kms$ and $40\,\kms$. The observational rates reported by \citet{li2020} (black markers) are much smaller than our estimated rates. However, if we recompute these observational rates using our default values for $\vel\tsc{eject}$ and $f\tsc{eject}$, they are in better agreement with our results (green markers).} 
    \label{fig:AccRate}
\end{figure}

Fig.\,\ref{fig:AccRate} shows the time evolution of the mean estimated accretion rate averaged over all lobes from all simulations, $\avg{\dot{M}_{\textsc{acc},l}}$, assuming $\vel\tsc{eject}\!=\!30\,\kms$ (orange line); the shaded orange region shows the mean rates obtained if $\vel\tsc{eject}$ is varied between $20\,\kms$ and $40\,\kms$. The blue line shows the evolution of the mean {\it true} accretion rate, defined as the total protostellar mass divided by the time since the protostar was born. We use this definition for {\it true} accretion rate as it is the definition used by \citet{li2020}. However, the actual accretion rate onto a protostar is highly episodic, as can be seen on Fig. 9 of \citep{Rohde18}. We attempt to estimate the episodic accretion rates in Section \ref{sec:bulletAcc}.

During Stage\,0, the estimated accretion rates fit the {\it true} accretion rates well, with a mean absolute error of $\mae{s0}{\avg{\dot{M}_{\textsc{acc},l}}}\!=\!1.2 \times 10^{-6}\,\Ms\,\yr^{-1}$, corresponding to a mean fractional error of $\maer{s0}{\avg{\dot{M}_{\textsc{acc},l}}}\!=\!0.23$. The estimated accretion rate becomes slightly less accurate during Stage\,I; there the mean absolute and fractional errors are $\mae{sI}{\avg{\dot{M}_{\textsc{acc},l}}}\!=\!1.5 \times 10^{-6}\,\Ms\,\yr^{-1}$ and $\maer{sI}{\avg{\dot{M}_{\textsc{acc},l}}}\!=\!0.39$. Overall the estimated accretion rates are accurate to a mean fractional error of $\maer{s0+I}{\avg{\dot{M}_{\textsc{acc},l}}}\!=\!0.36$.

The black markers represent the accretion rates estimated by \citet{li2020}. Their accretion rates are significantly lower than ours, but this is because they assume an ejection velocity of $\vel\tsc{eject}\!=\!500\,\kms$ and an ejection fraction of $f\tsc{eject}\!=\!0.3$. If we instead adopt our default values ($\vel\tsc{eject}\!=\!30\,\kms$ and $f\tsc{eject}\!=\!0.1$), the observational estimates (green markers) match ours much better.

%%%%%%%%%%%%%%%%%%%%%%%%%%%%%%%%%%%%%%%%%%%%%%%%%%
%%%%%%%%%%%%%%%%%%%%%%%%%%%%%%%%%%%%%%%%%%%%%%%%%%

%%%%%%%%%%%%%%%%%%%%%%%%%%%%%%%%%%%%%%%%%%%%%%%%%%
\begin{figure}
	\includegraphics[width=\columnwidth]{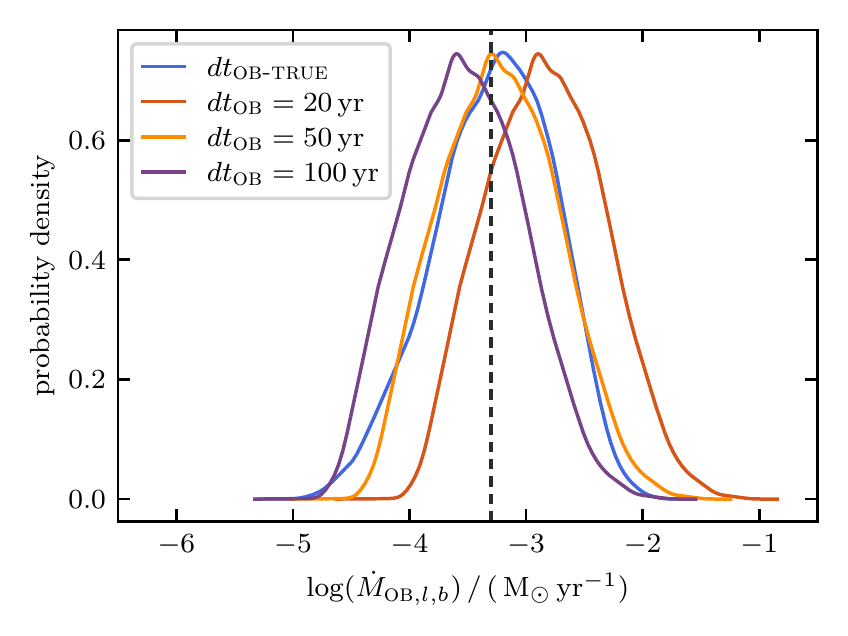}
    \caption{Probability density of the estimated accretion rate during an outburst event inferred from the properties of  the corresponding outflow bullet (see Eq.\,\ref{eq:accRateOB_1}). The vertical grey dashed line shows the true accretion rate, which is fixed at $\dot{M}\tsc{ob}\!=\!5 \times 10^{-4}  \,\Ms \, \mrm{yr}^{-1}$ (see Section \ref{sec:FeedbackModels}). The blue line shows the accretion rate estimated using the true outburst duration, $dt\tsc{ob-true}$, whilst the red, orange and purple lines show the accretion rates estimated using -- respectively -- $dt\tsc{ob}\!=\!20\,\rm{yr}$, $50\,\rm{yr}$ and $100\,\rm{yr}$. Accretion rates estimated using $dt\tsc{ob-true}$ and $dt\tsc{ob}\!=\!50\,\rm{yr}$ yield distributions peaked very close to the true rate, $\dot{M}\tsc{ob}$. Using $dt\tsc{ob}\!=\!20\,\rm{yr}$ or $100\,\rm{yr}$ respectively over- or under-estimates the accretion rate.} 
    \label{fig:bulletAcc}
\end{figure}
%%%%%%%%%%%%%%%%%%%%%%%%%%%%%%%%%%%%%%%%%%%%%%%%%%

%-------------------------------------------------
\subsubsection{Outburst accretion rate}
\label{sec:bulletAcc}
%-------------------------------------------------

Accretion onto young protostars is observed to be episodic rather than continuous. Outflow bullets are a consequence of these episodic accretion events. In the sequel we estimate the accretion rates during outbursts from the dynamics of the resulting outflow bullets, and compare these estimates with the actual accretion rates.

We compute the accretion rate required to trigger the simultaneous ejection of two oppositely directed bullets using an equation similar to Eq.\,\ref{eq:accRate}, viz.
\begin{eqnarray}\label{eq:accRateOB_1}
\dot{M}_{\textsc{ob},l,b}\!\!&\!\!=\!\!&\!\!\frac{2\,M_{l,b}\,\vel_{l,b}}{f\tsc{eject}\,\vel_{\textsc{eject}}\,dt\tsc{ob}}\\\nonumber
\!\!&\!\!=\!\!&\!\!8 \times 10^{-4}\,\Ms\,\yr^{-1}\,
\left( \frac{M_{l,b}}{0.003 \, \Ms} \right)
\left(\frac{\vel_{l,b}}{20 \, \kms} \right) \,
\hspace{1.2cm}\\\label{eq:accRateBulletInsert}
\!\!&\!\!\!\!&\!\!\hspace{0.95cm}\times\,
\left( \frac{f\tsc{eject}}{0.1} \right)^{-1} \,
\left( \frac{\vel\tsc{eject}}{30 \, \kms} \right)^{-1} \,
\left( \frac{dt\tsc{ob}}{50\,\yr} \right)^{-1};
\end{eqnarray}
the only differences from Eq.\,\ref{eq:accRate} are that we have replaced the mass of the lobe, $M_l$, with the mass of the bullet, $M_{l,b}$; the velocity of the lobe, $\vel_l$, with the velocity of the bullet, $\vel_{l,b}$; and the lifetime of the lobe, $\tau_l$, with the duration of the outburst $dt\tsc{ob}$.

Fig.\,\ref{fig:bulletAcc} compares the KDEs of the estimated and actual accretion rates during the outburst events. The true accretion rate of the sub-grid model during an outburst is always $\dot{M}\tsc{ob, true}\!=\!5.0 \times 10^{-4} \, \Ms \, \mrm{yr}^{-1}$ \citep[vertical black dashed line; see Section\,\ref{sec:FeedbackModels} and][]{Stamatellos12}. We estimate outburst accretion rates assuming $dt\tsc{ob}\!=\!20\,\rm{yr}$, $50\,\rm{yr}$ and $100\,\rm{yr}$, as shown by the red, orange and purple lines on Fig.\,\ref{fig:bulletAcc}; and also using the true outburst duration, as shown by the blue line on Fig.\,\ref{fig:bulletAcc}. In all cases we assume an ejection velocity of $\vel\tsc{eject}\!=\!30\, \kms$.

The distribution of accretion rates obtained assuming the true outburst duration (blue line on Fig.\,\ref{fig:bulletAcc}) has a mean value $\dot{M}\tsc{ob}\!=\!5.3\,\times\,10^{-4}\,\Ms\,\rm{yr}^{-1}$, very close to the true accretion rate of $\dot{M}\tsc{ob}\!=\!5.0\,\times\,10^{-4}\,\Ms\,\mrm{yr}^{-1}$. Similarly, the distribution of accretion rates obtained with $dt\tsc{ob}\!=\!50\,\rm{yr}$ (orange line on Fig.\,\ref{fig:bulletAcc}) has a mean value $\dot{M}\tsc{ob}\,=\,5.9\,\times\,10^{-4}\,\Ms\,\rm{yr}^{-1}$, again very close to the true accretion rate. The distributions obtained with $dt\tsc{ob}\!=\!20\,\rm{yr}$ and $100\,\rm{yr}$ have mean values of $\dot{M}\tsc{ob}\!=\!1.4 \times 10^{-3}\,\Ms\,\rm{yr}^{-1}$ and $\dot{M}\tsc{ob}\!=\!3 \times 10^{-4} \, \Ms \, \mrm{yr}^{-1}$, which are still reasonable estimates. We conclude that the dynamical information carried by the outflow bullets allows us to estimate the accretion rate during the corresponding accretion event.

%We perform least square fits of log-normal functions to the probability density distributions. According to the log-normal fit, the distribution for the true $dt\tsc{ob}$ peaks at $\dot{M}\tsc{ob}\,=\,4.2\,\times\,10^{-4} \, \Ms \, \mrm{yr}^{-1}$, close to the true accretion rate of $\dot{M}\tsc{ob}\,=\,5.0\,\times\,10^{-4} \, \Ms \, \mrm{yr}^{-1}$, with a width of $\sigma\,=\,0.89$. Similarly, the distribution for  $dt\tsc{ob}\,=\,50$ yr peaks at $\dot{M}\tsc{ob}\,=\,4.0\,\times\,10^{-4} \, \Ms \, \mrm{yr}^{-1}$ with a width of $\sigma\,=\,0.91$. The distributions for $dt\tsc{ob}\,=\,20$ and $100$ yr peak at $\dot{M}\tsc{ob}\,=\,7.2 \times 10^{-4}$ and $\dot{M}\tsc{ob}\,=\,2.5 \times 10^{-4} \, \Ms \, \mrm{yr}^{-1}$, still giving reasonable estimates. Therefore, the dynamical information carried by the outflow bullets allows us to estimate the accretion rate during the corresponding accretion event.

%-------------------------------------------------
\subsubsection{Time dependent accretion history}
\label{sec:accHistory}
%-------------------------------------------------

We can combine the estimated outflow-bullet ages from Section \,\ref{sec:tDynBullet}, with the estimated outburst accretion rates from Section\,\ref{sec:bulletAcc}, to reconstruct the accretion history of the underlying protostar between the current time $t_0$ and a time in the past $t_p\!=\! t_0 - \tau_{\textsc{max, bullet},l}$. Here, $\tau_{\textsc{max, bullet},l}$ (Eq. \ref{eq:dyn-bullet}) is the longest dynamical age among the bullets in the lobe with $\vel_{\textsc{max},l,b}\!>\!20 \, \kms$; typically it is $\la100\,\kyr$.

The total estimated accretion rate is the sum of the estimated episodic accretion rate, $\dot{M}_{\textsc{ob},l,b}$, which is only active during an outburst event, and a much lower estimated background accretion rate,  $\dot{M}_{\textsc{ob},l,b}$, i.e.
\begin{eqnarray}
\label{eq:AccRateHist}
\dot{M}_{\star,l}&=&\dot{M}_{\textsc{bg},l} + \dot{M}_{\textsc{ob},l,b}\,.
\end{eqnarray} 
The background accretion rate is given by 
\begin{eqnarray}
\label{eq:AccMassBG}
\dot{M}_{\textsc{bg},l}&=&\mathrm{max} \left(\frac{M_{\textsc{cont},l} - M_{\textsc{episodic},l}}{t_0 - t_p} , 0 \right).
\end{eqnarray}
In Eq.\,\ref{eq:AccMassBG}, $M_{\textsc{cont},l}$ is an estimate of all the mass accreted between $t_p$ and $t_0$, i.e. 
\begin{eqnarray}
\label{eq:AccMassCont}
M_{\textsc{cont},l}&=&(t_0 - t_p) \, \dot{M}_{\textsc{acc},l}\,, 
\end{eqnarray}
with $\dot{M}_{\textsc{acc},l}$ given by Eq.~\ref{eq:accRate}; and $M_{\textsc{episodic},l}$ is an estimate of the mass accreted during all the episodic accretion/outburst events during the same period,
\begin{eqnarray}
\label{eq:AccMassEpisodic}
M_{\textsc{episodic},l}&=&\sum_{b=0}^{n_{\textsc{bullet},l}} dt_{\textsc{ob}} \, \dot{M}_{\textsc{ob},l,b}  \, ,
\end{eqnarray}
with $n_{\textsc{bullet},l}$ the number of distinct bullets, $dt_{\textsc{ob}}\!=\!50\,\rm{yr}$, and $\dot{M}_{\textsc{ob},l,b}$ from Eq. \ref{eq:accRateOB_1}. %We note that the estimated background accretion rate can become negative if $M_{\textsc{cont},l} < M_{\textsc{episodic},l}$, in which case we cannot make an estimate.%%not needed, now max in eq 44

\begin{figure}
	\includegraphics[width=\columnwidth]{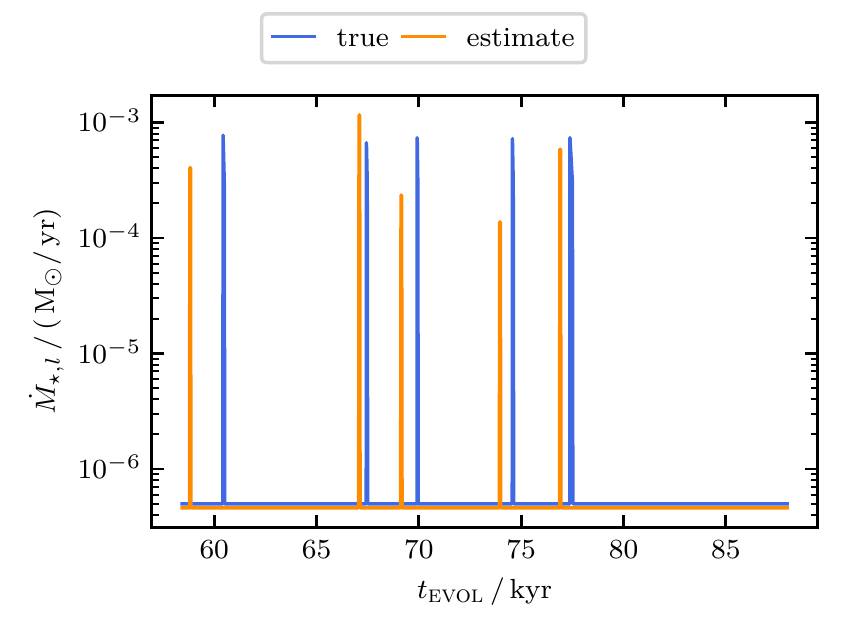}
    \caption{The accretion history for simulation \sn{5}{0.5}{2}{10}. The blue line shows the true accretion rate, and the orange line shows the accretion rate reconstructed using estimated outflow-bullet ages from Section \,\ref{sec:tDynBullet}, and estimated outburst accretion rates from Section\,\ref{sec:bulletAcc}.} 
    \label{fig:AccHist_single}
\end{figure}

The orange line on Fig.\,\ref{fig:AccHist_single} shows the accretion history of simulation \sn{5}{0.5}{2}{10} reconstructed in this way, and the blue line shows the true accretion history, demonstrating that the dynamical properties of outflow bullets allow us to reconstruct the accretion history well. The background accretion rate and the number of outburst events are reproduced almost exactly. The timings of the outburst events are accurate to $\lesssim1\,\kyr$, and the accretion rates during outburst to within a factor $\lesssim2$.

%%%%%%%%%%%%%%%%%%%%%%%%%%%%%%%%%%%%%%%%%%%%%%%%%%
%%%%%%%%%%%%%%%%%%%%%%%%%%%%%%%%%%%%%%%%%%%%%%%%%%

%-------------------------------------------------
\section{Caveats}
\label{sec:Caveats}
%-------------------------------------------------

In this work, we have used simulations to evaluate the accuracy of some methods frequently used to estimate protostellar ages, accretion histories and ouitflow rates. Here, we discuss the most important limitations of this study.

Cores have a range of masses, described by the core mass function \citep{Konyves20}, whereas our simulations only treat $1\,\Ms$ cores, which limits the generality of our results. However, the final protostellar mass function of the simulations is compatible with the observed protostellar mass function \citep{Rohde20}. Therefore, our results should approximately represent what happens in a low-mass star-forming region.

We perform hydrodynamic simulations and do not include magnetic fields. The importance of magnetic fields in launching outflows is implicit in our sub-grid outflow model. Magnetic fields are also expected to enhance the stability of accretion discs \citep[e.g.][]{Wurster19a}, but our episodic accretion model regulates the stability of the inner disc and therefore limits the influence of magnetic fields there \citep{Lomax16b,Lomax16a}. However, magnetic fields might slow down the collapse of our cores, and might also influence the propagation of the outflows.

Protostellar jets are observed to be embedded in low-velocity, wide-angle winds. Since these wide-angle winds are launched from radii far out in the accretion disc, between $\sim\!10\,\rm{AU}$ and $\sim\!50\,\rm{AU}$ \citep{Louvet18, Podio21, Pascucci20, Lee21}, they might play an essential role in removing angular momentum from these regions, thus allowing gas to spiral inwards faster \citep{Lee20rev}. Our sub-grid outflow model episodically injects wide-angle winds but does not account for the associated removal of angular momentum. A single episodic wide-angle wind, similar to that in our outflow model, has been observed by \citep{Zhang19}, and further observations are needed to establish how common such winds are. However, self-consistently treated wide-angle winds are probably dynamically not very important due to their low velocities of about $10\,\kms$. 

Although the radiative feedback model treats the star's close surrounding, including the inner accretion disc, it does not account for the effect of radiation on the outflow cavity \citep{Stamatellos07, Stamatellos11}. Since we are simulating low-mass star formation, we expect the effect of full radiative transfer to be rather limited. For example, the additional momentum due to radiation pressure
\begin{eqnarray}
p\tsc{rad}&=&\sum \limits_{i}^{n_{\star}} \sum \limits_{t}^{n\tsc{step}}\,\left\{\frac{L_{\star,i}}{c\,dt}\right\}
\end{eqnarray}
(where $n\tsc{step}$ is the number of simulation time steps, $dt$, and $c$ is the speed of light), would add $\sim\!1\%$ to the total outflow momentum. Thus with radiative feedback the outflow cavities would widen somewhat faster than in our simulations \citep{Kuiper16}.

The actual mechanism underlying outflow launching is not well understood. Many different aspects could alter how the gas is ejected. Our simulations are limited to one specific model of outflow launching with well-defined parameters, for example the ejection fraction. This parametrisation is motivated by the study of \citet{Matzner99}, who show that at sufficiently large distances from the driving source, all hydro-magnetic winds behave similarly. Moreover, as shown by \citet{Rohde18}, protostellar outflows are self-regulated, in the sense that varying the sub-grid model parameters has a limited impact on the outflow properties. Therefore, we believe that our simulated outflows should be evolving similarly to those of FUor-like stars.

%-------------------------------------------------
\section{Conclusions}
\label{sec:Conclusions}
%-------------------------------------------------
The early phases of stellar evolution and gas accretion are closely linked to, and regulated by, the launching of protostellar outflows. Therefore, these outflows carry fossil information about the stellar age and accretion history. Outflows extend far into the protostellar environment and are relatively easy to observe and resolve. We compute outflow properties from a set of 44 SPH simulations with episodic outflow feedback, estimate stellar ages and accretion histories from the outflows, and compare these estimates with the underlying simulations. In this way, we compute the uncertainties inherent in different observational methods for estimating ages and outflow properties -- but ignoring intrinsic observational uncertainties and selection effects.

To compute the outflow properties, we extract the outflow lobes from our simulations and use the \textsc{Optics} clustering algorithm to trace individual outflow bullets. Here we summarise our results, demonstrating that protostellar outflows are indeed a useful "window on the past".
\begin{itemize}
\item{Assuming momentum conservation, we estimate the gas entrainment factor $\epsilon$, i.e. the ratio of outflowing mass to ejected mass. The outflowing mass is much larger than the ejected mass, because gas in the parental core is swept up by the outflow. During Stage\,0 we find good agreement with the true entertainment factor of $\epsilon\!\sim\!10$. During Stage\,I, the estimated entrainment factor is larger than the true entrainment factor.}
\item{We compute dynamical ages for the outflows using five different methods and compare them. Estimating the dynamical age from the outflow front is very accurate during Stage\,0, but becomes increasingly inaccurate during Stage I. Conversely, the perpendicular method is not very reliable during Stage\,0, but very accurate during Stage\,I, with a fractional error of $\sim\!15\%$. The most commonly used method for deriving the dynamical age is based on taking the ratio of the outflow length to the highest outflow velocity, but it has a fractional error of $\sim\!34\%$. We propose a new method to estimate the dynamical age from two successive outflow bullets: if two distinct bullets are present in an outflow cavity, this method provides good age estimates for both Stage\,0 and Stage\,I, with a fractional error of $\sim\!24\%$. These errors are for the case of nearly perfect information, observed dynamical ages probably come with significanty larger errors.}
\item{We find that dynamical ages of individual outflow bullets accurately describe their true age, especially if the bullets are young and have not yet swept up a significant amount of extra mass. On average, we over-predict the ages of bullets by $\sim\!0.44\,\kyr$, but for recently ejected bullets by only $\sim\!0.07\,\kyr$.}
\item{We find that outflow cavities widen over time, as observed. We fit an opening angle--age relation, similar to the one derived by \citet{Arce06}, but conclude that estimating the stellar age using this relation is not advisable, due to the large uncertainties caused by both the large variation in opening angles, and the shallow slope of the relation. However, we find that the opening angle can be used to differentiate between Stage\,0 and Stage\,I sources. From our simulations, only $10\,\%$ of all lobes have opening angles smaller than $\sim \,65^{\circ}$, and only $10\,\%$ have opening angles larger than $\sim\!100^{\circ}$, when entering Stage\,I.}
\item{Using these dynamical ages, as well as the true protostellar ages, we estimate the outflow rates of mass, momentum and energy and compare them. We find that these rates are rather constant during Stage\,0, and diminish slightly once the protostars enter Stage\,I. This reduction in outflow activity can help to distinguish between Stage\,0 and Stage\,I sources.}
%With an uncertainty of 10 \% our outflow launching sources with accretion rates $ > 10^{-5} \, \Ms \, \yr^{-1}$ can be classified as Stage\,0 and with accretion rates $ < 10^{-6} \, \Ms \, \yr^{-1}$ as Stage\,I.
\item{The estimated outflow rates and entrainment factors allow us to reconstruct the protostellar accretion rates. During Stage\,0 these estimates are on average accurate to $\sim\,23\,\%$; during Stage\,I the estimate is less accurate.}
\item{Using the derived dynamical properties of a bullet, we can estimate the accretion rate during the outburst event associated with the ejection of that bullet. The distribution of the estimated accretion rates peaks around the true accretion rate. These accretion rates, together with the bullet age, allow us to reconstruct rather accurately the protostellar accretion history.}
\end{itemize}

Overall, protostellar outflows carry useful information which can be used to estimate the protostellar age and evolutionary Stage. Focusing on individual outflow bullets reveals the episodic accretion behaviour and allows the reconstruction of the accretion history.

%-------------------------------------------------
\section*{acknowledgements}
%-------------------------------------------------
The authors like to thank the anonymous referee for comments that helped to significantly improve the paper.
PFR, SW  and SDC acknowledge support via the ERC starting grant No.
679852 `RADFEEDBACK'. 
DS and SW thank the DFG for funding
via the SFB 956 `Conditions \& impact of star formation', via the sub-projects C5 and C6.
APW gratefully acknowledges the support of a consolidated
grant (ST/K00926/1) from the UK Science and Technology Facilities Council. 
The authors gratefully acknowledge the Gauss Centre for Supercomputing e.V. 
(www.gauss-centre.eu) for funding this project (ID: pr47pi) by providing computing time on the GCS Supercomputer SuperMUC at Leibniz Supercomputing Centre (www.lrz.de).
PR acknowledges D. Price for providing the visualisation tool SPLASH \citep{Price11}.

%-------------------------------------------------
\section*{data availability}
%-------------------------------------------------
The data underlying this article will be shared on reasonable request to the corresponding author.



\bibliographystyle{mnras}
\bibliography{Bib} % if your bibtex file is called example.bib

%\begin{thebibliography}{99}
%\end{thebibliography}

%%%%%%%%%%%%%%%%%%%%%%%%%%%%%%%%%%%%%%%%%%%%%%%%%%
\appendix
%%%%%%%%%%%%%%%%% APPENDICES %%%%%%%%%%%%%%%%%%%%%

\section{Inclination}
\label{app:Inclination}
\begin{table}
\caption{Correction factors for various outflow properties depending on the inclination angle $\theta$ (i.e. the angle between the outflow direction and the line of sight). Assuming that all orientations have an equal probability, the mean inclination is $\theta\,=\,57.3^{\circ}$ \citep{Bontemps96}. This table is based on that in \citet{Li19}.} 
\begin{center}\begin{tabular}{ccccc}
\hline 
Outflow  & Inclination  & \multicolumn{3}{c}{Correction angle} \\
 parameter & dependence & ${\theta}\,=\,57.3^{\circ}$ & ${\theta}\,=\,5^{\circ}$ & ${\theta}\,=\,85^{\circ}$ \\
\hline
$\tau_{l}$ & $ \mrm{cos}(\theta) / \mrm{sin}(\theta)$ & 0.6 & 11.4 & 0.09 \\
$l_{l}$ & $1 / \mrm{sin}(\theta)$ & 1.2 & 11.5 & 1.0 \\
$\vel_{l}$ & $1 / \mrm{cos}(\theta)$ & 1.9 & 1.0 & 11.5 \\
$p_{l}$ & $1 / \mrm{cos}(\theta)$ & 1.9 & 1.0 & 11.5 \\
$\dot{M}_{\textsc{out},l}$ & $\mrm{sin}(\theta) / \mrm{cos}(\theta)$ & 1.7 & 0.09 & 11.4 \\
$F_{l}$ & $\mrm{sin}(\theta) / \mrm{cos}^2(\theta)$ & 2.9 & 0.09 & 131.2 \\
$L_{l}$ & $\mrm{sin}(\theta) / \mrm{cos}^3(\theta)$ & 5.3 & 0.09 & 1505 \\
\end{tabular}\end{center} 
\label{Table:incl}
\end{table}
\FloatBarrier

%%%%%%%%%%%%%%%%%%%%%%%%%%%%%%%%%%%%%%%%%%%%%%%%%%
\section{Outflow launching}
\label{app:outflow_launching}

Assuming protostellar outflows are launched magneto-centrifugally as described by the x-wind model \citep{Shu94}, \citet{Shu95} show that the almost radially ejected winds become collimated to a cylindrical configuration
\begin{eqnarray}
\label{eq:shu95}
    \rho\tsc{w} \propto 1\,/\,(r\,\mathrm{sin}(\theta))^2 \, .
\end{eqnarray}
Here, $\rho\tsc{w}$ is the density of the wind and $\theta$ the angle of the flow with respect to the outflow axis. \citet{Matzner99} generalize this finding to any momentum conserving MHD wind in an environment with a power-law density distribution. For radially ejected winds, it follows from Eq. \ref{eq:shu95} that 
\begin{eqnarray}
\label{eq:Matzner1}
    \rho\tsc{w} \vel\tsc{w}^2 \propto 1\,/\,(r\,\mathrm{sin}(\theta))^2 \, ,
\end{eqnarray}  
where $\vel\tsc{w}$ is the wind velocity. \citet{Matzner99} argue that an angular force distribution, $P(\mu)$, must be flat for $\mu \rightarrow 0$ and therefore approximate the distribution with
\begin{eqnarray}
\label{eq:Matzner2}
P(\theta)&\propto&r^2\rho\tsc{w}\,\upsilon\tsc{w}^2\;\,\simeq\;\,\left[\ln\left(\frac{2}{\theta_{\textsc{jet}}}\right) \left(\sin^2(\theta)+\theta_{\textsc{jet}}^2\right)\right]^{-1}\!.\hspace{0.6cm}
\end{eqnarray} 
where $\theta_{\textsc{jet}}$ is the angular scale over which the distribution is flattened.
This force distribution for MHD winds at large distances from the launching point is often called the "wind-driven shell" model and is used by numerous authors to approximate outflows in sub-grid outflow implementations \citep[e.g.][]{Cunningham11, Offner14, Kuiper15, Tanaka18, Rohde18, Grudi20}.
Similarly, \citet{Rohde18} assume this force distribution (Eq. \ref{eq:Matzner2}) and separate it into distinct density and velocity distributions that satisfy the force distribution
\begin{eqnarray}\label{Eq:Mrho}
\rho_{\textsc{inject}}(\theta)&\propto&P^{1/2}(\theta)\,,
\end{eqnarray}
\begin{eqnarray}\label{Eq:Mvel}
|\upsilon_{\textsc{inject}}(\theta)|&\propto&P^{1/4}(\theta)\,.
\end{eqnarray} 
The free parameter $\theta_{\textsc{jet}}$, regulating the collimation of the outflow, and $\theta_{\textsc{open}}$, the opening angle where we cut-off the force distribution, have a rather limited influence on the outcome of the simulations \citep{Rohde18}. Here, we use the default parameters from \citet{Rohde18}, $\theta_{\textsc{jet}}\,=\,0.01$ and  $\theta_{\textsc{open}}\,=\,0.4$. Due to the cut-off at $\theta_{\textsc{open}}\,=\,0.4$ the mean value of the velocity distribution (Eq. \ref{Eq:Mvel}) is $\sim 2.2$.

%%%%%%%%%%%%%%%%%%%%%%%%%%%%%%%%%%%%%%%%%%%%%%%%%%

%%%%%%%%%%%%%%%%%%%%%%%%%%%%%%%%%%%%%%%%%%%%%%%%%%

% Don't change these lines % Why?
\bsp	% typesetting comment
\label{lastpage}
\end{document}